\documentclass[preprint,12pt]{elsarticle}
\journal{International Journal of Engineering Sciences}
\usepackage{graphics,epsfig}
\usepackage{graphicx}
\usepackage{epstopdf}
\usepackage{float}
\usepackage{amssymb,amstext,amsmath}
\usepackage{mathtools}
\mathtoolsset{showonlyrefs}
\usepackage{booktabs}
\usepackage{natbib}
\begin{document}
\begin{frontmatter}
\title{An asymptotically exact theory of functionally graded piezoelectric shells}
\author{K. C. Le\footnote{Phone: +49 234 32-26033, email: chau.le@rub.de.}}
\address{Lehrstuhl f\"{u}r Mechanik - Materialtheorie, Ruhr-Universit\"{a}t Bochum,\\D-44780 Bochum, Germany}
\begin{abstract} 
An asymptotically exact two-dimensional theory of functionally graded piezoelectric shells is derived by the variational-asymptotic method. The error estimation of the constructed theory is given in the energetic norm. As an application, analytical solution to the problem of forced vibration of a functionally graded piezoceramic cylindrical shell with thickness polarization fully covered by electrodes and excited by a harmonic voltage is found. 
\end{abstract}

\begin{keyword}
piezoelectric, functionally graded, shell, variational-asymptotic method.
\end{keyword}

\end{frontmatter}

\section{Introduction}

Functionally graded materials (FGM) were first invented by a group of Japanese scientists  \citep{niino1990recent,koizumi1992recent} and have been since then widely used in smart structures for the active vibration control \citep{he2001active,jha2013critical}. Such smart structures in form of plates or shells are quite often made of the functionally graded piezoelectric (FGP) materials whose electroelastic moduli vary through the thickness \citep{wu1996piezoelectric}. If oscillating voltages, as external excitations, are controlled on electrodes covering faces or boundaries of such structures, then, under certain conditions, the structures may exhibit anti-resonant regime that can be used to eliminate unwanted vibrations \citep{preumont2002vibration}. Mention that, from the formal mathematical point of view, smart sandwich structures with piezo patches bonded to elastic layers \citep{bailey1985distributed,crawley1987use,crawley1991induced,tzou1989theoretical} also belong to the FGP-structures with piecewise constant electroelastic moduli. 

Due to the above mentioned inhomogeneous material properties of FGP-structures, the problems of their equilibrium and vibration admit exact analytical solutions of the three-dimensional theory of piezoelectricity only in a few exceptional cases (see, e.g., \citep{zhong2003three,vel2004three,pan2005exact,elishakoff2015mechanics} and the references therein). By this reason different approaches have been developed depending on the type of the structures. If FGP-plates and shells are thick, no accurate two-dimensional  theory can be constructed, so only the numerical methods or semi-analytical methods applied to three-dimensional theory of piezoelectricity make sense \citep{li2008three,dong2008three,wen2011three}. However, if FGP-plates and shells are thin, the reduction from the three- to two-dimensional theory is possible and different approximations can be constructed. Up to now two main approaches have been developed: (i) the variational approach based on Hamilton's variational principle and on some ad-hoc assumptions generalizing Kirchhoff-Love's hypothesis to FGP-plates and shells \citep{he2001active,han2001transient,wu2002high,yiqi2010nonlinear}\footnote{The literature on this topic is huge due to the variety of the 2-D FGP-shell and plate theories: single-layer, multi-layer, refined theories including rotary inertias and transverse shears et cetera. It is therefore impossible to cite all references. For the overview the reader may consult \citep{reddy2004mechanics,liew2011review,shen2016functionally} and the references therein.}, (ii) the asymptotic approach based on the analysis of the three-dimensional equations of piezoelectricity, mainly for the laminated FGP-plates \citep{cheng2000cthree,wu2009cylindrical,leugering2012asymptotic}. The disadvantage of the variational approach is the necessity of having an Ansatz for the displacements and electric field that is difficult to be justified, while simplicity and brevity are its advantages. The asymptotic method needs no a priori assumptions; however, the direct asymptotic analysis of the 3-D differential equations of piezoelectricity is very cumbersome. The synthesis of these two approaches, called the variational-asymptotic method, first proposed by \citet{berdichevsky1979variational} and developed further by \citet{le1999vibrations}, seems to avoid the disadvantages of both approaches described above and proved to be quite effective in constructing approximate equations for thin-walled structures. Note that this method has been applied, among others, to derive the 2-D theory of homogeneous piezoelectric shells by \citet{le1984fundamental,Le86a}, the 2-D static theory of purely elastic sandwich plates and shells by \citet{berdichevsky2010nonlinear,berdichevsky2010asymptotic}, the theory of smart beams by \citet{roy2007asymptotically}, the theory of low- and high frequency vibration of laminate composite shells by \citet{lee2009adynamic,lee2009bdynamic}, and just recently, the theory of smart sandwich shells by \citet{le2016asymptotically}. Note also the closely related method of  gamma convergence used in homogenization of periodic and random microstructures \citep{braides2002gamma} and dimension reduction of plate theories \citep{friesecke2006hierarchy}.

The aim of this paper is to construct the rigorous first order approximate two-dimensional FGP-shell theory by the variational-asymptotic method. We consider the FGP-shell whose electroelastic moduli vary in the thickness direction. The dimension reduction is based on the asymptotic analysis of the action functional containing small parameters that enables one to find the distribution of the displacements and electric field from the solution of the so-called thickness problem. Using the generalized Prager-Synge identity for the inhomogeneous piezoelectric body, we provide also the error estimation of the constructed theory in the energetic norm. We apply this theory to the problem of forced vibration of a functionally graded piezoceramic cylindrical shell with thickness polarization fully covered by electrodes and excited by a harmonic voltage. The exact analytical solution to this problem is found.

The paper is organized as follows. After this short introduction the variational formulation of the problem is given in Sections 2 and 3. Sections 4,5 are devoted to the asymptotic analysis of the action functional. In Section 6 the two-dimensional theory of FGP-shells is obtained. Section 7 provides the error estimation of the constructed theory. Section 8 presents the exact analytical solution to the problem of forced vibration of a circular cylindrical FGP-shell. Finally, Section 9 concludes the  paper.

\section{Variational principle of piezoelectricity}

Let $\mathcal{V}\subseteq \mathbb{R}^3$ be a domain of the three-dimensional euclidean space occupied by a linear and inhomogeneous piezoelectric body in its stress-free undeformed state. A motion of this body is completely determined by two fields, namely, the displacement  field ${\bf w}(\mathbf{x},t)$, and the electric potential $\varphi (\mathbf{x},t)$. For simplicity we will consider the case of purely electrical loading on the body, which corresponds to specifying the value of the electric potential on the electrodes. Let the boundary of the body, $\partial \mathcal{V}$, be decomposed into $n+1$ two-dimensional surfaces $\mathcal{S}_e^{(1)}, \ldots , \mathcal{S}_e^{(n)}$, and $\mathcal{S}_d$. The subboundaries $\mathcal{S}_e^{(1)},\ldots , \mathcal{S}_e^{(n)}$ are covered by electrodes. We assume that the electrodes are infinitely thin so that their kinetic and electroelastic energies can be neglected compared with those of the body. On these electrodes the electric potential is prescribed 
\begin{equation}
\varphi =\varphi _{(i)}(t) \quad \text{on
$\mathcal{S}_e^{(i)},i=1,\ldots ,n$}.
\label{2.1}
\end{equation}
Hamilton's variational principle of piezoelectricity (see, e.g., \citep{Le86a,le1999vibrations}) states that the true displacement $\check{\mathbf{w}}(\mathbf{x},t)$ and electric potential $\check{\varphi }(\mathbf{x},t)$ of an inhomogeneous piezoelectric body change in space and time in such a way that the action functional
\begin{equation}
\label{2.2}
I[\mathbf{w}(\mathbf{x},t),\varphi (\mathbf{x},t)]=\int_{t_0}^{t_1}\int_{\mathcal{V}}[T(\mathbf{x},\dot{\mathbf{w}})-W(\mathbf{x},\boldsymbol{\varepsilon} ,\mathbf{E})]\, dv \, dt
\end{equation} 
becomes stationary among all continuously differentiable functions $\mathbf{w}(\mathbf{x},t)$ and $\varphi (\mathbf{x},t)$ satisfying the initial and end conditions
\begin{displaymath}
\mathbf{w}(\mathbf{x},t_0)=\mathbf{w}_0(\mathbf{x}), \quad \mathbf{w}(\mathbf{x},t_1)=\mathbf{w}_1(\mathbf{x}),
\end{displaymath}
as well as constraints \eqref{2.1}. The integrand in the action functional \eqref{2.2} is called Lagrangian, while $dv$ is the volume element and the dot over quantities denotes the partial time derivative. In the Lagrangian $T(\mathbf{x},\dot{\mathbf{w}})$ describes the kinetic energy density given by\footnote{As we shall be concerned with mechanical vibrations of non-conducting piezoelectric bodies at frequencies far below optical frequencies, the coupling between the electric and magnetic fields and the dependence of the kinetic energy on $\dot{\varphi }$ can be neglected.}
\begin{equation}
\label{2.3}
T(\mathbf{x},\dot{\mathbf{w}})=\frac{1}{2}\rho (\mathbf{x}) \dot{\mathbf{w}} \cdot \dot{\mathbf{w}},
\end{equation}
with $\rho (\mathbf{x})$ being the mass density. Function $W(\mathbf{x},\boldsymbol{\varepsilon} ,\mathbf{E})$, called electric enthalpy density, reads 
\begin{equation}
\label{2.4}
W(\mathbf{x},\boldsymbol{\varepsilon} ,\mathbf{E})=\frac{1}{2}\boldsymbol{\varepsilon} \mathbf{:}\mathbf{c}_E(\mathbf{x})\mathbf{:}\boldsymbol{\varepsilon}-\mathbf{E}\cdot \mathbf{e}(\mathbf{x})\mathbf{:}\boldsymbol{\varepsilon}-\frac{1}{2}\mathbf{E}\cdot \boldsymbol{\epsilon}_S(\mathbf{x})\cdot \mathbf{E},
\end{equation}
where $\boldsymbol{\varepsilon} $ is the strain tensor
\begin{equation}
\label{2.4a}
\boldsymbol{\varepsilon }=\frac{1}{2}(\nabla \mathbf{w}+(\nabla \mathbf{w})^T),
\end{equation}
while $\mathbf{E}$ the electric field
\begin{equation}
\label{2.4b}
\mathbf{E}=-\nabla \varphi .
\end{equation}
Applying the standard calculus of variation one easily shows that the stationarity condition $\delta I=0$ implies the equations of motion of piezoelectric body (including the equation of electrostatics)
\begin{equation}
\label{2.5}
\rho (\mathbf{x})\ddot{\mathbf{w}}=\text{div}\boldsymbol{\sigma }, \quad \text{div}\mathbf{D}=0,
\end{equation}
where the stress tensor field $\boldsymbol{\sigma }$ and the electric induction field $\mathbf{D}$ are given by
\begin{equation}
\label{2.6}
\begin{split}
\boldsymbol{\sigma }=\frac{\partial W}{\partial \boldsymbol{\varepsilon}}=\mathbf{c}_E(\mathbf{x})\mathbf{:}\boldsymbol{\varepsilon}-\mathbf{E}\cdot \mathbf{e}(\mathbf{x}),
\\
\mathbf{D}=-\frac{\partial W}{\partial \mathbf{E}}=\mathbf{e}(\mathbf{x})\mathbf{:}\boldsymbol{\varepsilon}+\boldsymbol{\epsilon}_S(\mathbf{x})\cdot \mathbf{E}.
\end{split}
\end{equation}
We call $\mathbf{c}_E(\mathbf{x})$ the (fourth-rank) tensor of elastic stiffnesses, $\mathbf{e}(\mathbf{x})$ the (third-rank) tensor of piezoelectric constants, while $\boldsymbol{\epsilon}_S(\mathbf{x})$ the (second-rank) tensor of dielectric permittivities.\footnote{The label $E$ in $\mathbf{c}_E(\mathbf{x})$ indicates elastic stiffnesses at constant electric field, while the label $S$ in $\boldsymbol{\epsilon}_S(\mathbf{x})$ denotes dielectric permittivities at constant strain for the piezoelectric material.} For the elastic (dielectric) material $\mathbf{e}(\mathbf{x})=0$, so it is the degenerate case of piezoelectric material. Substituting the constitutive equations \eqref{2.6} into \eqref{2.5} and making use of the kinematic equations \eqref{2.4a} and \eqref{2.4b}, we get the closed system of four governing equations for four unknown functions $w_i(\mathbf{x},t)$ and $\varphi (\mathbf{x},t)$. These equations are subjected to the following boundary conditions
\begin{equation}
\label{2.6a}
\boldsymbol{\sigma }\cdot \mathbf{n}=0 \quad \text{on $\partial \mathcal{V}$}, \quad
\mathbf{D}\cdot \mathbf{n}=0 \quad \text{on $\mathcal{S}_d$},
\end{equation}
and \eqref{2.1}, where $\mathbf{n}$ is the unit outward normal vector to the boundary.

\section{Variational formulation for FGP-shells}
\begin{figure}[htb]
	\centering
	\includegraphics[width=7cm]{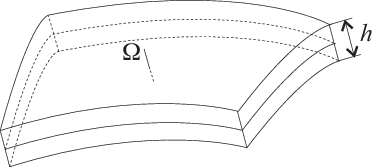}
	\caption{A portion of a shell}
	\label{fig:1}
\end{figure}

Let $\Omega $ be a two-dimensional smooth surface bounded by a smooth closed curve $\partial \Omega $. At each point of the surface $\Omega $ a  segment of length $h$ in the direction perpendicular to the surface is drawn so that its center lies on the surface. If the length $h$ is sufficiently small, the segments do not intersect each other and fill the domain $\mathcal{V}$ occupied by a shell in its undeformed state shown schematically in Fig.~\ref{fig:1}, with $\Omega $ being the shell middle surface. 
Thus, the undeformed shell  occupies the domain specified by the equation
\begin{displaymath}
z^i(x^\alpha ,x)=r^i(x^\alpha )+xn^i(x^\alpha ),
\end{displaymath}
where $z^i=r^i(x^\alpha )$ is the equation of the middle surface $\Omega $, and $n^i(x^\alpha )$ are the cartesian components of the normal vector ${\bf n}$ to this surface. We shall use Latin indices, running from 1 to 3, to refer to the spatial co-ordinates and the Greek indices, running from 1 to 2, to refer to the surface co-ordinates $x^1$ and $x^2$. The curvilinear co-ordinates $x^\alpha $ take values in a domain of $\mathbb{R}^2$, while $x\in [-h/2,h/2]$. 
We analyze the forced vibration of the shell made of a functionally graded piezoelectric material. Let $\Omega _\pm $ denote the face surfaces of the shell corresponding to $x=\pm h/2$ in the above equation. We consider three methods of electrode arrangement encountered most often:
\begin{itemize}
\item[(i)] There are no electrodes on the face surfaces of the shell (unelectroded face surfaces). The edge of the shell is partially electroded (see Figure 
\ref{fig:2}).
\begin{figure}[htb]
    \begin{center}
    \includegraphics[height=4cm]{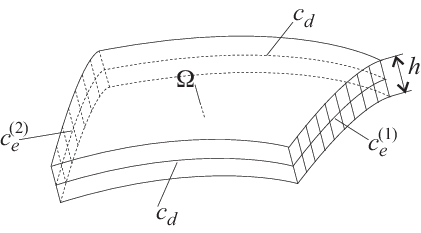}
    \end{center}
    \caption{Partially electroded edge of a piezoelectric shell.}
    \label{fig:2}
\end{figure}
Thus, we assume that the contour $\partial \Omega $ is decomposed into open curves $c_e^{(1)},\ldots ,c_e^{(n)}$ (where there are electrodes) and the remaining part $c_d$. For $x^a\in c_e^{(i)}\times [-h/2,h/2]$ on the electroded portions of the edge the electric potential is prescribed
\begin{equation}
\varphi =\varphi _{(i)}(t), \quad i=1,\ldots ,n.
\label{3.1}
\end{equation}
On the unelectroded portion of the boundary the electric charge should vanish.
\item[(ii)] The face surfaces $\Omega _\pm $ are fully coated by the electrodes. They form two equipotential surfaces where
\begin{equation}
\varphi =\pm \varphi _0(t)/2, \quad \text{for $x=\pm h/2$}.
\label{3.2}
\end{equation}
The difference between these values, $\varphi _0(t)$, is called voltage for short.
\item[(iii)] The face surfaces of the shell are only partially coated by electrodes on $\mathcal{S}_{e\pm} \subset \Omega _\pm $. The remaining face surfaces $\mathcal{S}_{d\pm}$ are uncoated. This case can be regarded as the mixed situation of the two cases above.
\end{itemize}

For the asymptotic analysis of the FGP-shell inhomogeneous in the normal direction it is convenient to use the curvilinear coordinates $\{ x^\alpha ,x\}$ introduced above and the co- and contravariant index notation for vectors and tensors, with Einstein's summation convention being employed. In this coordinate system the action functional reads 
\begin{equation*}
I=\int_{t_0}^{t_1}\int_{\Omega}\int_{-h/2}^{h/2}[T(x,\dot{\mathbf{w}})-W(x,\boldsymbol{\varepsilon} ,\mathbf{E})] \kappa \, dx\, da\, dt,
\end{equation*}
where $\kappa =1-2Hx+Kx^2$ (with $H$ and $K$ being the mean and Gaussian curvature of the middle surface, respectively) and $da$ denotes the area element of the middle surface. The explicit dependence of the Lagrangian on the transverse coordinate $x$ of this functionally graded material is precisely indicated. The kinetic energy density becomes 
\begin{equation*}
T(x,\dot{\mathbf{w}})=\frac{1}{2}\rho (x) (a^{\alpha \beta }\dot{w}_\alpha \dot{w}_\alpha +\dot{w}^2),
\end{equation*}
where $a^{\alpha \beta }$ are the contravariant components of the surface metric tensor, while $w_\alpha $ and $w$ are the projections of the displacement vector onto the tangential and normal directions to the middle surface
\[
w_\alpha =t^i_\alpha w_i=r^i_{,\alpha }w_i,\quad w=n^iw_i.
\]
The electric enthalpy density $W$ reads
\begin{equation*}
W(x,\boldsymbol{\varepsilon} ,\mathbf{E})=\frac{1}{2}c^{abcd}_E(x)\varepsilon _{ab}\varepsilon _{cd} -e^{cab}(x)\varepsilon _{ab}E_c -\frac{1}{2}\epsilon ^{ab}_S(x)E_a E_b.
\end{equation*}

The problem is to replace the three-dimensional action functional by an approximate two-dimensional action functional for a thin FGP-shell, whose functions depend only on the longitudinal co-ordinates $x^1,x^2$ and time $t$. The possibility of reduction of the three- to the two-dimensional functional is related to the smallness of the ratios between the thickness $h$ and the characteristic radius of curvature $R$ of the shell middle surface and between $h$ and the characteristic scale of change of the electroelastic state in the longitudinal directions $l$ (see \citep{le1999vibrations} and Section 5). We assume that
\begin{displaymath}
\frac{h}{R}\ll 1, \quad \frac{h}{l}\ll 1.
\end{displaymath}
Additionally, we assume that 
\begin{equation}
\frac{h}{ c\tau }\ll 1,
\label{3.3}
\end{equation}
where $\tau $ is the characteristic scale of change of the functions $w_i$ and $\varphi $ in time (see \citep{le1999vibrations}) and $c$ the minimal velocity of plane waves in the piezoelectric materials under consideration. This means that we consider in this paper only statics or low-frequency vibrations of the functionally graded piezoelectric shell. By using the variational-asymptotic method, the two-dimensional action functional will be constructed below in which terms of the order $h/R$ and $h/l$ are neglected as compared with unity (the first-order or ``classical'' approximation).

In order to fix the domain of the transverse co-ordinate in the passage to the limit $h\to 0$, we introduce the dimensionless co-ordinate
\begin{equation*}
\zeta =\frac{x}{h}, \quad \zeta \in [-1/2,1/2],
\end{equation*}
and transform the action functional to
\begin{equation}\label{3.4}
I=\int_{t_0}^{t_1}\int_{\Omega}\int_{-1/2}^{1/2}h[T(\zeta ,\dot{\mathbf{w}})-W(\zeta ,\boldsymbol{\varepsilon} ,\mathbf{E})] \kappa \, d\zeta \, da\, dt,
\end{equation}
Now $h$ enters the action functional explicitly through the components of the strain tensor $\varepsilon _{ab}$ and the electric field $E_a$ 
\begin{align}
\varepsilon _{\alpha \beta }&=
w_{(\alpha ;\beta )}-b_{\alpha \beta }w-
h\zeta b^\lambda _{(\alpha }w_{\lambda ;\beta )}+h\zeta 
c_{\alpha \beta }w,
\notag \\
2\varepsilon _{\alpha 3}&= 
\frac{1}{h}w_{\alpha |\zeta }+w_{,\alpha }+b^\lambda _\alpha w_\lambda 
-\zeta b^\lambda _\alpha w_{\lambda |\zeta }, \quad \varepsilon _{33}=\frac{1}{h}w_{|\zeta },
\label{3.5} \\
E_\alpha &=-\varphi _{,\alpha }, \quad E_3=-\frac{1}{h}\varphi 
_{|\zeta }.
\notag
\end{align}
Here and below, the semicolon preceding Greek indices denotes the covariant derivatives on the surface, while the parentheses surrounding a pair of indices the symmetrization operation. Note that the raising or lowering of indices of surface tensors will be done with the surface metrics $a^{\alpha \beta }$ and $a_{\alpha \beta }$, $b_{\alpha \beta }$ and $c_{\alpha \beta }$ are the second and third fundamental forms of the surface, vertical bar followed by $\zeta $ indicates the partial derivative with respect to $\zeta $ and {\it not} with respect to $x^\zeta $. 

\section{Two-dimensional electro-elastic moduli}

Before applying the variational-asymptotic procedure to functional \eqref{3.4} let us transform the electric enthalpy density to another form more convenient for the asymptotic analysis \citep{Le86a}. We note that, among terms of $W(\varepsilon _{ab},E_a)$, the derivatives $w_{\alpha |\zeta }/h$ and $w_{|\zeta }/h$ in $\varepsilon _{\alpha 3}$ and $\varepsilon _{33}$ as well as $E_3=-\varphi _{|\zeta }/h$ are the main ones in the asymptotic sense. Therefore it is convenient to single out the components $\varepsilon _{\alpha 3}$ and $\varepsilon _{33}$ as well as $E_3$ in the electric enthalpy density. We represent the latter as the sum of two quadratic forms $W_\parallel $ and $W_\perp $ corresponding to longitudinal and transverse electric enthalpy densities, respectively. These are defined by
\begin{align}
W_\parallel &=\min_{\varepsilon _{\alpha 3},
\varepsilon _{33}}\max_{E_3}W, \notag
\\
W_\perp &=W-W_\parallel . \label{4.1}
\end{align}
Let us first find the decomposition \eqref{4.1} in the most general case of anisotropy \citep{Le86a}. Long, but otherwise simple calculations show that
\begin{align}
W_\parallel &=\frac{1}{2}c^{\alpha \beta \gamma \delta }_N(\zeta )
\varepsilon _{\alpha \beta }\varepsilon _{\gamma \delta }-
e^{\gamma \alpha \beta }_N(\zeta )\varepsilon _{\alpha \beta }E_\gamma 
-\frac{1}{2}\epsilon^{\alpha \beta }_N(\zeta ) E_\alpha E_\beta ,
\notag \\
W_\perp &=\frac{1}{2}c^{3333}_E(\zeta )\gamma ^2+c^{\alpha 333}_E(\zeta )\gamma
\gamma _\alpha +\frac{1}{2}c^{3\alpha 3\beta }_E(\zeta )\gamma _\alpha
\gamma _\beta 
\label{4.2} \\
&-e^{333}(\zeta )\gamma F-e^{3\alpha 3}(\zeta )\gamma _\alpha F
-\frac{1}{2}\epsilon^{33}_S(\zeta ) F^2,
\notag
\end{align}
where
\begin{align*}
\gamma &=\varepsilon _{33}+r^{\alpha \beta }(\zeta )\varepsilon _{\alpha \beta 
}-r^\alpha (\zeta ) E_\alpha ,
\\
\gamma _\alpha &=2\varepsilon _{\alpha 3}+p_\alpha ^{\mu \nu }(\zeta )
\varepsilon _{\mu \nu }-p_\alpha ^\mu (\zeta ) E_\mu ,
\\
F&=E_3+q^{\alpha \beta }(\zeta )\varepsilon _{\alpha \beta }+q^\alpha (\zeta )
E_\alpha .
\end{align*}
Functions $c^{\alpha \beta \gamma \delta }_N(\zeta )$, $e^{\gamma \alpha \beta }_N(\zeta )$, $\epsilon^{\alpha \beta }_N(\zeta )$, $c^{3\alpha 3\beta }_E(\zeta )$, $c^{\alpha 333}_E(\zeta )$, $c^{3333}_E(\zeta )$, $\epsilon^{33}_S(\zeta )$, $e^{333}(\zeta )$, $e^{3\alpha 3}(\zeta )$, $r^{\alpha \beta }(\zeta )$, $r^\alpha (\zeta )$, $p_\alpha ^{\mu \nu }(\zeta )$, $p_\alpha ^\mu (\zeta )$, $q^{\alpha \beta }(\zeta )$, and $q^\alpha (\zeta )$ can be regarded as components of surface tensors referred to the basis vectors $\mathbf{t}_\alpha $ of the middle surface. We shall call them ``two-dimensional'' electroelastic moduli. They are evaluated in terms of the three-dimensional moduli by means of the formulas
\begin{gather}
c^{\alpha \beta \gamma \delta }_N=c^{\alpha \beta \gamma \delta 
}_P+q^{\alpha \beta }e^{3\gamma \delta }_P, \quad e^{\gamma
\alpha \beta }_N=e^{\gamma \alpha \beta }_P-q^{\alpha \beta }
\epsilon^{\gamma 3}_P,
\notag \\
\epsilon^{\alpha \beta }_N=\epsilon^{\alpha \beta }_P-
q^\alpha \epsilon^{\beta 3}_P, \quad q^{\alpha \beta }=
e^{3\alpha \beta }_P/\epsilon^{33}_P, \quad q^\alpha =
\epsilon^{\alpha 3}_P/\epsilon^{33}_P,
\notag \\
c^{\alpha \beta \gamma \delta }_P=\bar{c}^{\alpha \beta \gamma \delta }
-k^{\alpha \beta }_\nu \bar{c}^{\gamma \delta \nu 3}, \quad
e^{a\alpha \beta }_P=\bar{e}^{a\alpha \beta }-k^{\alpha \beta }_\nu 
\bar{e}^{a\nu 3},
\notag \\
\epsilon^{\alpha b}_P=\bar{\epsilon}^{\alpha b}+k^\alpha _\nu 
\bar{e}^{b\nu 3},\quad \epsilon^{33}_P=\bar{\epsilon}^{33}+k_\nu
\bar{e}^{3\nu 3},
\notag 
\\
k^{\mu \nu }_\alpha =h_{\alpha \beta }\bar{c}^{\mu \nu \beta 3},\quad
k^\mu _\alpha =h_{\alpha \beta }\bar{e}^{\mu \beta 3},\quad 
k_\alpha =h_{\alpha \beta }\bar{e}^{3\beta 3},\quad 
h_{\alpha \beta }=(\bar{c}^{3\alpha 3\beta })^{-1},
\label{4.3} 
\\
\bar{c}^{a\alpha b\beta }=c^{a\alpha b\beta }_E-c^{a\alpha 33}_E
c^{b\beta 33}_E/c^{3333}_E, \quad \bar{e}^{ab\beta }=e^{ab\beta }
-c^{b\beta 33}_Ee^{a33}/c^{3333}_E,
\notag 
\\
\bar{\epsilon}^{ab}=\epsilon^{ab}+e^{a33}e^{b33}/
c^{3333}_E,\quad p^{\mu \nu }_\alpha =k^{\mu \nu }_\alpha +
k_\alpha q^{\mu \nu },
\notag 
\\
p^\mu _\alpha =k^\mu _\alpha -k_\alpha q^\mu ,\quad 
r^{\alpha \beta } =f^{\alpha \beta }+fq^{\alpha \beta },\quad
r^{\alpha } =f^{\alpha }-fq_{\alpha },
\notag 
\\
f^{\alpha \beta }=\frac{c^{\alpha \beta 33}_E-c^{\lambda 333}_Ek^{\alpha
\beta }_\lambda }{c^{3333}_E},\quad f^\alpha =\frac{e^{\alpha 33}
-c^{\lambda 333}_Ek^\alpha _\lambda }{c^{3333}_E},\quad 
f =\frac{e^{333}-c^{\lambda 333}_Ek_\lambda }{c^{3333}_E},
\notag
\end{gather}
where their dependence on the variable $\zeta $ is suppressed for short.  

Note that, as these tensors are referred to the basis $\{ \mathbf{t}_\alpha , \mathbf{n}\}$, their components will depend on $\zeta $ through the shifter $\mu ^\beta _\alpha =\delta ^\beta _\alpha -h\zeta b^\beta _\alpha $ even for homogeneous shells. However, it can be shown for the FGP-shell inhomogeneous in the normal direction that components of two-dimensional electroelastic moduli of any rank, denoted symbolically by $A(\zeta )$, possess the property
\[
A(\zeta )=A_0(\zeta )+O(\frac{h}{R}) A_0(\zeta ),
\]
where $A_0(\zeta )$ describe the {\it physical} inhomogeneity of the material and do not depend on the shifter (as in the case of FGP-plates with cartesian coordinates), while the factor $O(h/R)$ in the second term is due to the shifter $\mu ^\beta _\alpha $ of the curvilinear coordinates solely. Therefore, when constructing 2-D shell theories having the error $h/R$ as compared with unity, it can be assumed that $A(\zeta )=A_0(\zeta )$.

Let us note certain special symmetry cases.
\begin{itemize}
\item {\it Mirror planes parallel to the middle surface}. 
If properties of the piezoelectric material are invariant under
reflections relative to planes parallel to the middle surface, then 
the following 2-D tensors vanish
\[
c^{\alpha 333}_E=0,\, e^{333}=0,\quad p^{\mu \nu }_\alpha =0,\,
p^\mu _\alpha =0,\quad q^{\mu \nu }=0,\, q^\mu =0,
\]
and
\[
c^{\alpha \beta \gamma \delta }_N=c^{\alpha \beta \gamma \delta }_P,
\quad e^{\gamma \alpha \beta }_N=e^{\gamma \alpha \beta }_P,\quad
\varepsilon ^{\alpha \beta }_N=\varepsilon ^{\alpha \beta }_P.
\]
\item {\it n-fold rotation axes that coincide with
the normal to the middle surface}. When $n$ 
is even, all 2-D tensors of odd rank vanish
\[
e^{\gamma \alpha \beta }_N=0,\, c^{\alpha 333}_E=0,\, e^{3\alpha 3}=0,\,
r^\alpha =0,\, p^{\mu \nu }_\alpha =0,\, q^\mu =0.
\]
\item {\it Transverse isotropy}. When\index{Transverse isotropy} 
properties of
the piezoelectric material are invariant under rotations about
the normal to the middle surface (model of a piezoceramic shell
polarized along the normal with symmetry $\infty \cdot m$), it can be
shown that all 2-D tensors of odd rank vanish; the tensor 
$c^{\alpha \beta \gamma \delta }_N$ has the form
\[
c^{\alpha \beta \gamma \delta }_N=c^N_1a^{\alpha \beta }a^{\gamma \delta }
+c^N_2(a^{\alpha \gamma }a^{\beta \delta }+a^{\alpha \delta }a^{\beta 
\gamma }),
\]
and all the 2-D tensors of second rank are spherical.
\end{itemize}

\section{\bf Asymptotic analysis of the action functional}

We restrict ourselves to the low frequency vibrations of the FGP-shell for which assumption \eqref{3.3} is valid. Based on this assumption we may neglect the kinetic energy density in the variational-asymptotic procedure.\footnote{For the high-frequency vibrations of elastic and piezoelectric shells and rods where the kinetic energy density should be kept in the variational-asymptotic analysis see \citep{berdichevsky1980high,Berdichevsky82,le1982high,Le85b,Le86b,le1997high,le1999vibrations,lee2009bdynamic}.}  Since there are two different cases of unelectroded and electroded faces of the shell, we shall do the asymptotic analysis for these cases separately.

\subsection{Unelectroded faces.}
We could start the variational-asymptotic procedure with the determination of the set $\mathcal{N}$ according to its general scheme \citep{le1999vibrations}. As a result, it would turn out that, at the first step, the functions ${\bf w}$ and $\varphi $ do not depend on the transverse
co-ordinate $\zeta $: ${\bf w}={\bf u}(x^\alpha ,t)$, $\varphi =\psi (x^\alpha ,t)$; at the second step the function ${\bf w}^\star $ is a linear function of $\zeta $; and at the next step ${\bf w}^{\star \star }$ and $\varphi ^{\star \star }$ are completely determined through ${\bf u}$ and $\psi $. Thus, the set $\mathcal{N}$ according to the variational-asymptotic scheme consists of functions ${\bf u}(x^\alpha ,t)$ and $\psi (x^\alpha ,t)$. We will pass over these long, but otherwise standard, deliberations and make a change of unknown functions immediately.

We introduce the following functions
\begin{equation}
\begin{split}
u(x^\alpha ,t)=\langle w(x^\alpha ,\zeta ,t) \rangle ,\quad u_\alpha (x^\alpha ,t)=\langle w_\alpha (x^\alpha ,\zeta ,t)\rangle ,
\\
\psi (x^\alpha ,t)=\langle \varphi (x^\alpha ,\zeta ,t) \rangle ,
\end{split}
\label{5.1}
\end{equation}
where $\langle . \rangle $ denotes the integration over $\zeta $ within the limits $[-1/2,1/2]$.
Functions $u, u_\alpha $ correspond to the mean displacements of the shell, while $\psi $ describes the mean electric potential. Now let us make the following change of unknown functions 
\begin{align}
w(x^\alpha ,\zeta ,t)&=u(x^\alpha ,t)+hy (x^\alpha ,\zeta ,t),
\notag \\
w_\alpha (x^\alpha ,\zeta ,t)&=u_\alpha (x^\alpha ,t)-h\zeta \varphi _{\alpha } 
(x^\alpha ,t) +hy_\alpha (x^\alpha ,\zeta ,t),
\label{5.2} \\
\varphi (x^\alpha ,\zeta ,t)&=\psi (x^\alpha ,t)+h\chi (x^\alpha ,\zeta ,t),
\notag
\end{align}
where
\begin{equation}
\varphi _\alpha =u_{,\alpha }+b^\beta _\alpha u_\beta .
\label{5.3}
\end{equation}
Because of definitions \eqref{5.1} functions $y$, $y_\alpha $, and $\chi $ should satisfy the following constraints
\begin{equation}
\langle y \rangle =0, \quad \langle y_\alpha \rangle =0, 
\quad
\langle \chi \rangle =0.
\label{5.4}
\end{equation}
Equations \eqref{5.2} and \eqref{5.4} set up a one-to-one correspondence between $w, w_\alpha ,\varphi $ and the set of functions $u, u_\alpha ,\psi ,y,y_\alpha ,\chi $ and determine the change in the unknown functions $\{ w,w_\alpha ,\varphi \} \to \{
u,u_\alpha ,\psi ,y,y_\alpha ,\chi \}$.

Asymptotic analysis enables one to determine the order of smallness of $y,y_\alpha ,\chi $. If these terms are neglected, then \eqref{5.2} is a generalization of the well-known Kirchhoff-Love
hypotheses to a piezoelectric shell. The electroelastic state of a shell is then characterized by the measures of extension $A_{\alpha \beta }=u_{(\alpha ;\beta )} -b_{\alpha \beta } u$, the measures of bending $B_{\alpha \beta }=u_{;\alpha \beta }
+(u_\lambda b^\lambda _{(\alpha })_{;\beta )}
+b^\lambda _{(\alpha }u_{\lambda ;\beta )}-c_{\alpha \beta }u $ and, finally, the surface electric field $F_\alpha =-\psi _{,\alpha }$. We introduce the following notation
\begin{gather*}
\varepsilon _A=\max_{\mathcal{S}}\sqrt{A_{\alpha \beta 
}A^{\alpha \beta }}, \quad \varepsilon _B=h\max_{\mathcal{S}}
\sqrt{B_{\alpha \beta }B^{\alpha \beta }},\quad
f_F=\max_{\mathcal{S}}\sqrt{F_\alpha F^\alpha },
\\
\Delta _\alpha =\max_{\mathcal{B}}|y_{\alpha |\zeta }|,\quad
\Delta =\max_{\mathcal{B}}|y_{|\zeta }|,\quad
\Pi =\max_{\mathcal{B}}|\chi _{|\zeta }|.
\end{gather*}
Consider a certain point of the middle surface $\Omega $. The best constant $l$ in the inequalities
\begin{equation*}
\begin{split}
\left| A_{\alpha \beta ,\gamma }\right| \le \frac{\varepsilon _A}{l}, \quad h\left| B_{\alpha \beta ,\gamma }\right| 
\le \frac{\varepsilon _B}{l},\quad \left| F_{\alpha ,\beta }
\right| \le \frac{f_F}{l}, 
\\
\max_\zeta \left| y_{\alpha ,\beta }\right| \le \frac{\Delta _\alpha}{l} 
,  \quad \max_\zeta \left| y_{,\alpha }\right| 
\le \frac{\Delta }{l}, \quad \max_\zeta \left| \chi _{,\alpha }\right| 
\le \frac{\Pi }{l}
\end{split}
\end{equation*}
is called the characteristic scale of change of the electroelastic state in the longitudinal directions. We define the inner domain $\Omega _i$ as a subdomain of $\Omega$ in which the following inequalities hold:
\begin{equation}
h/R\ll 1,\quad h/l\ll 1.
\label{5.5}
\end{equation}
We assume the domain $\Omega $ to consist of the inner domain $\Omega _i$ and a domain $\Omega _b$ abutting on the contour $\partial \Omega$ with width of the order $h$ (boundary layer). Then functional \eqref{3.4} can be decomposed into the sum of two functionals, an inner one for which an iteration process will be applied, and a boundary layer functional. As in the theory of elastic shells, the boundary layer functional can be neglected in the first-order
approximation. Therefore, the problem reduces to finding stationary points of the inner functional that can be identified with the functional \eqref{3.4} ($\Omega _i\equiv \Omega$).

We now fix $u,u_\alpha ,\psi $ and seek $y,y_\alpha ,\chi $. Substituting \eqref{5.2} into the action functional \eqref{3.4}, we will keep in it the asymptotically principal terms containing $y,y_\alpha ,\chi $ and neglect all other terms. The estimations based on the above inequalities lead to the asymptotic formulas
\begin{equation}
\varepsilon _{\alpha \beta }=A_{\alpha \beta }- h\zeta B_{
\alpha \beta },\quad 2\varepsilon _{\alpha 3}=
y_{\alpha |\zeta },\quad \varepsilon _{33}=y_{|\zeta }.
\label{5.6}
\end{equation}
It is also easy to check that, within the first-order approximation,
\begin{equation}
E_\alpha =F_\alpha ,\quad E_3=-\chi _{|\zeta }.
\label{5.7}
\end{equation}
According to formulas \eqref{5.6} and \eqref{5.7} the longitudinal electric enthalpy does not contain asymptotically principal terms containing $y,y_\alpha ,\chi $ and can be neglected. Since the transverse electric enthalpy contains only the derivatives of $y,y_\alpha ,\chi $ with respect to $\zeta $, we drop the integration over $\Omega $ and $t$ and reduce the thickness problem to finding extremal of the functional
\begin{multline}
I_\perp =\frac{h}{2} \int _{-1/2}^{1/2} 
[c^{3333}_E(\zeta )\gamma ^2+2c^{\alpha 333}_E(\zeta )\gamma
\gamma _\alpha +c^{3\alpha 3\beta }_E(\zeta )\gamma _\alpha
\gamma _\beta 
\\
-2e^{333}(\zeta )\gamma F-2e^{3\alpha 3}(\zeta )\gamma _\alpha F
-\epsilon ^{33}_S(\zeta ) F^2 ]\, d\zeta ,
\label{5.8}
\end{multline}
where
\begin{align*}
\gamma &=y_{|\zeta }+r^{\alpha \beta }(\zeta )(A_{\alpha \beta 
}-h\zeta B_{\alpha \beta })-r^\alpha (\zeta )F_\alpha ,
\\
\gamma _\alpha &=y_{\alpha |\zeta }+p_{\alpha }^{\mu \nu }(\zeta )
(A_{\mu \nu }-h\zeta B_{\mu \nu })-p_\alpha ^\mu (\zeta )F_\mu ,
\\
F&=-\chi _{|\zeta }+q^{\alpha \beta }(\zeta )(A_{\alpha \beta 
}-h\zeta B_{\alpha \beta })+q^\alpha (\zeta )F_\alpha .
\end{align*}
We minimize functional \eqref{5.8} in $y,y_\alpha $ and maximize in $\chi $ under constraints \eqref{5.4}. Obviously, the minimax value of $I_\perp $ is equal to zero and is attained at
$\gamma =\gamma _\alpha =F=0$. Integrating the equations
\begin{align}
y_{|\zeta }&=-r^{\alpha \beta }(\zeta )(A_{\alpha \beta 
}-h\zeta B_{\alpha \beta })+r^\alpha (\zeta )F_\alpha , \notag
\\
y_{\alpha |\zeta }&=-p_{\alpha }^{\mu \nu }(\zeta )
(A_{\mu \nu }-h\zeta B_{\mu \nu })+p_\alpha ^\mu (\zeta )F_\mu ,\label{5.8a}
\\
\chi _{|\zeta }&=q^{\alpha \beta }(\zeta )(A_{\alpha \beta 
}-h\zeta B_{\alpha \beta })+q^\alpha (\zeta )F_\alpha \notag
\end{align}
that follow from $\gamma =\gamma _\alpha =F=0$ and taking constraints \eqref{5.4} into account, we obtain
\begin{align}
y(x^\alpha ,\zeta ,t)&=-A_{\alpha \beta } \mathcal{I}[r^{\alpha \beta }]+F_\alpha \mathcal{I}[r^\alpha ] 
+hB_{\alpha \beta } \mathcal{I}[r^{\alpha \beta }\zeta ] ,
\notag \\
y_\alpha (x^\alpha ,\zeta ,t)&=-A_{\mu \nu } \mathcal{I}[p_{\alpha }^{\mu \nu }]+F_\mu \mathcal{I}[p_\alpha ^\mu ]
 +hB_{\mu \nu } \mathcal{I}[p_{\alpha }^{\mu \nu } \zeta ] ,
\label{5.9} \\
\chi (x^\alpha ,\zeta ,t)&=A_{\alpha \beta } \mathcal{I}[q^{\alpha \beta }]+F_\alpha \mathcal{I}[q^\alpha ] 
-hB_{\alpha \beta } \mathcal{I}[q^{\alpha \beta } \zeta ].
\end{align}
Here and below, for any function $f(\zeta )$, $\mathcal{I}[f]$ is the function of $\zeta $ defined by
\begin{equation}
\label{5.10}
\mathcal{I}[f](\zeta )=\int_{-1/2}^\zeta f(\xi )\, d\xi -\langle \int_{-1/2}^\zeta f(\xi )\, d\xi \rangle .
\end{equation}
Thus, according to this definition $\langle \mathcal{I}[f](\zeta )\rangle =0$, and functions $y,y_\alpha ,\chi $ from \eqref{5.9} fulfill constraints \eqref{5.4}.

\subsection{Electroded faces.}
In this case the electric potential $\varphi $ should obey constraints \eqref{3.2}. Consequently, we make another change of the unknown functions:
\begin{align}
w(x^\alpha ,\zeta ,t) &=u(x^\alpha ,t)+hy (x^\alpha ,\zeta ,t),
\notag \\
w_\alpha (x^\alpha ,\zeta ,t) &=u_\alpha (x^\alpha ,t)-h\zeta \varphi _{\alpha } 
(x^\alpha ,t) +hy_\alpha (x^\alpha ,\zeta ,t),
\label{5.11} \\
\varphi (x^\alpha ,\zeta ,t) &=\varphi _0 (t)\zeta +h\chi (x^\alpha ,\zeta ,t),
\notag
\end{align}
where $\varphi _\alpha $ is given by \eqref{5.3}. Thus, the difference between \eqref{5.2} and \eqref{5.11} concerns only the first term of $\varphi $, where $\varphi _0\zeta $ is
substituted in place of $\psi $. In view of \eqref{3.2} we impose constraints
\begin{equation}
\langle y \rangle =0,\quad \langle y_\alpha \rangle =0,
\quad
\chi |_{\zeta =\pm 1/2} =0
\label{5.12}
\end{equation}
on functions $y,y_\alpha ,\chi $.

Let us introduce the following notation
\begin{gather*}
\varepsilon _A=\max_{\mathcal{S}}\sqrt{A_{\alpha \beta 
}A^{\alpha \beta }}, \quad \varepsilon _B=h\max_{\mathcal{S}}
\sqrt{B_{\alpha \beta }B^{\alpha \beta }},
\\
\Delta _\alpha =\max_{\mathcal{B}}|y_{\alpha |\zeta }|,\quad
\Delta =\max_{\mathcal{B}}|y_{|\zeta }|,\quad
\Pi =\max_{\mathcal{B}}|\chi _{|\zeta }|.
\end{gather*}
We define the characteristic scale of change of the electroelastic state in the longitudinal directions as the best constant $l$ in the inequalities
\begin{equation*}
\begin{split}
\left| A_{\alpha \beta ,\gamma }\right| \le \frac{\varepsilon _A}{l}, \quad h\left| B_{\alpha \beta ,\gamma }\right| \le \frac{\varepsilon _B}{l},
\\
\max_\zeta \left| y_{\alpha ,\beta }\right| \le \frac{\Delta _\alpha }{l} 
,  \quad \max_\zeta \left| y_{,\alpha }\right| 
\le \frac{\Delta }{l}, \quad \max_\zeta \left| \chi _{,\alpha }\right| 
\le \frac{\Pi }{l},
\end{split}
\end{equation*}
and make the same assumption as in \eqref{5.5}. An estimation procedure analogous to the previous case leads to the following asymptotic formulas
\begin{equation}
\begin{split}
\varepsilon _{\alpha \beta }=A_{\alpha \beta }- h\zeta B_{
\alpha \beta },\quad 2\varepsilon _{\alpha 3}=
y_{\alpha |\zeta },\quad \varepsilon _{33}=y_{|\zeta },
\\
E_\alpha =0 ,\quad E_3=-\frac{\varphi _0}{h}-\chi _{|\zeta }
\end{split}
\label{5.13}
\end{equation}
that hold true within the first-order approximation. Fixing $u,u_\alpha $ and substituting \eqref{5.11} into the action functional \eqref{3.4}, we keep the asymptotically principal terms containing $y,y_\alpha ,\chi $. Since the obtained functional involves only the derivatives with respect to $\zeta $, we drop the integration over $\Omega $ and $t$ and reduce the thickness problem to finding extremal of the functional
\begin{align}
I_\perp &=\frac{h}{2} \int _{-1/2}^{1/2}  
[c^{3333}_E(\zeta )\gamma ^2+2c^{\alpha 333}_E(\zeta )\gamma
\gamma _\alpha +c^{3\alpha 3\beta }_E(\zeta )\gamma _\alpha
\gamma _\beta 
\notag \\
&-2e^{333}(\zeta )\gamma F-2e^{3\alpha 3}(\zeta )\gamma _\alpha F
-\epsilon ^{33}_S(\zeta ) F^2 ] \, d\zeta ,
\label{5.14}
\end{align}
under constraints \eqref{5.12}, where
\begin{align}
\gamma &=y_{|\zeta }+r^{\alpha \beta }(\zeta )(A_{\alpha \beta 
}-h\zeta B_{\alpha \beta }),\notag
\\
\gamma _\alpha &=y_{\alpha |\zeta }+p_{\alpha }^{\mu \nu }(\zeta )
(A_{\mu \nu }-h\zeta B_{\mu \nu }), \label{5.14a}
\\
F&=-\frac{\varphi _0}{h}-\chi _{|\zeta }+q^{\alpha \beta }(\zeta )(A_{\alpha \beta 
}-h\zeta B_{\alpha \beta }). \notag
\end{align}
Varying functional \eqref{5.14} with respect to $y,y_\alpha ,\chi $, we obtain the equations
\begin{equation}
\begin{split}
(c^{3333}_E\gamma +c^{\alpha 333}_E\gamma _\alpha -e^{333}F)_{|\zeta 
}=0,
\\
(c^{\alpha 333}_E\gamma +c^{3\alpha 3\beta }_E\gamma _\beta 
-e^{3\alpha 3}F)_{|\zeta }=0,
\\
(e^{333}\gamma +e^{3\alpha 3}\gamma _\alpha +\epsilon ^{33}_S 
F)_{|\zeta }=0,
\end{split}
\label{5.15}
\end{equation}
subjected to the boundary conditions at $\zeta =\pm 1/2$
\begin{equation}
\begin{split}
c^{3333}_E\gamma +c^{\alpha 333}_E\gamma _\alpha -e^{333}F=0,
\\
c^{\alpha 333}_E\gamma +c^{3\alpha 3\beta }_E\gamma _\beta 
-e^{3\alpha 3}F=0,
\\
\chi =0.
\end{split}
\label{5.16}
\end{equation}
Equations \eqref{5.15} and \eqref{5.16} yield
\begin{equation}
\begin{split}
c^{3333}_E\gamma +c^{\alpha 333}_E\gamma _\alpha -e^{333}F=0,
\\
c^{\alpha 333}_E\gamma +c^{3\alpha 3\beta }_E\gamma _\beta 
-e^{3\alpha 3}F=0,
\\
e^{333}\gamma +e^{3\alpha 3}\gamma _\alpha +\epsilon ^{33}_S 
F=D,
\end{split}
\label{5.17}
\end{equation}
where $D$ is independent of $\zeta $. Solving the first two equations of \eqref{5.17} with respect to $\gamma $ and $\gamma _\alpha $, we obtain
\[
\gamma =f(\zeta )F, \quad \gamma _\alpha =k_\alpha (\zeta )F, 
\]
with $f(\zeta )$ and $k_\alpha (\zeta )$ being taken from \eqref{4.3}. Substituting these formulas into the last equation and solving it with respect to $F$, we find that
\begin{equation}
\label{5.18}
F=\frac{D}{\epsilon ^{33}_P(\zeta )}.
\end{equation}
From this equation follows
\begin{equation}
\label{5.19}
\chi _{|\zeta }=-\frac{\varphi _0}{h}-\frac{D}{\epsilon ^{33}_P(\zeta )}+q^{\alpha \beta }(\zeta )(A_{\alpha \beta 
}-h\zeta B_{\alpha \beta }).
\end{equation}
As $\chi (-1/2)=0$, we integrate the above equation to get 
\begin{equation}
\label{5.20}
\chi (\zeta )=-\frac{\varphi _0}{h}(\zeta +\frac{1}{2})-\int_{-1/2}^\zeta \frac{Dd\xi }{\epsilon ^{33}_P(\xi )}+A_{\alpha \beta }\int_{-1/2}^\zeta q^{\alpha \beta }(\xi )d\xi -hB_{\alpha \beta } \int_{-1/2}^\zeta q^{\alpha \beta }(\xi )\xi d\xi .
\end{equation} 
Now, the constant $D$ can be found from the condition $\chi (1/2)=0$ giving
\begin{equation}
\label{5.21}
D=\langle \frac{1}{\epsilon_P^{33}(\zeta )} \rangle ^{-1}(-\frac{\varphi _0}{h}+A_{\alpha \beta }\langle q^{\alpha \beta }(\zeta )\rangle -hB_{\alpha \beta } \langle q^{\alpha \beta }(\zeta )\zeta \rangle ),
\end{equation}
and, consequently,
\begin{equation}
F=\frac{1}{\epsilon ^{33}_P(\zeta )}\langle \frac{1}{\epsilon_P^{33}} \rangle ^{-1}(-\frac{\varphi _0}{h}+A_{\alpha \beta }\langle q^{\alpha \beta }\rangle -hB_{\alpha \beta } \langle q^{\alpha \beta }\zeta \rangle ).
\label{5.22}
\end{equation}
We turn now to the equations for $y$ and $y_\alpha $. Substituting $F$ from above into the equations $\gamma =f(\zeta )F$ and $\gamma _\alpha =k_\alpha (\zeta )F$ and using \eqref{5.14a} we obtain for $y$
\begin{multline}
\label{5.23}
y_{|\zeta }=-\frac{f(\zeta )}{\epsilon ^{33}_P(\zeta )}\langle \frac{1}{\epsilon_P^{33}} \rangle ^{-1}\frac{\varphi _0}{h}-A_{\alpha \beta }(r^{\alpha \beta}(\zeta )-\frac{f(\zeta )}{\epsilon ^{33}_P(\zeta )}\langle \frac{1}{\epsilon_P^{33}} \rangle ^{-1}\langle q^{\alpha \beta }\rangle )
\\
+hB_{\alpha \beta }(r^{\alpha \beta}(\zeta )\zeta -\frac{f(\zeta )}{\epsilon ^{33}_P(\zeta )}\langle \frac{1}{\epsilon_P^{33}} \rangle ^{-1}\langle q^{\alpha \beta } \zeta \rangle ),
\end{multline}
and for $y_\alpha $
\begin{multline}
\label{5.24}
y_{\alpha |\zeta }=-\frac{k_\alpha (\zeta )}{\epsilon ^{33}_P(\zeta )}\langle \frac{1}{\epsilon_P^{33}} \rangle ^{-1}\frac{\varphi _0}{h}-A_{\mu \nu}(p_\alpha ^{\mu \nu }(\zeta )-\frac{k_\alpha (\zeta )}{\epsilon ^{33}_P(\zeta )}\langle \frac{1}{\epsilon_P^{33}} \rangle ^{-1}\langle q^{\mu \nu }\rangle )
\\
+hB_{\mu \nu }(p_\alpha ^{\mu \nu}(\zeta )\zeta -\frac{k_\alpha (\zeta )}{\epsilon ^{33}_P(\zeta )}\langle \frac{1}{\epsilon_P^{33}} \rangle ^{-1}\langle q^{\mu \nu } \zeta \rangle ).
\end{multline}
The integration of these equations taking into account the constraints \eqref{5.12} leads to 
\begin{multline}
\label{5.25}
y(x^\alpha ,\zeta ,t)=-\langle \frac{1}{\epsilon_P^{33}} \rangle ^{-1}\frac{\varphi _0}{h}\mathcal{I}[\frac{f}{\epsilon ^{33}_P}]-A_{\alpha \beta }\mathcal{I}[r^{\alpha \beta}-\frac{f}{\epsilon ^{33}_P}\langle \frac{1}{\epsilon_P^{33}} \rangle ^{-1}\langle q^{\alpha \beta }\rangle )]
\\
+hB_{\alpha \beta }\mathcal{I}[r^{\alpha \beta} \zeta -\frac{f}{\epsilon ^{33}_P}\langle \frac{1}{\epsilon_P^{33}} \rangle ^{-1}\langle q^{\alpha \beta } \zeta \rangle ],
\end{multline}
and
\begin{multline}
\label{5.26}
y_{\alpha }(x^\alpha ,\zeta ,t)=-\langle \frac{1}{\epsilon_P^{33}} \rangle ^{-1}\frac{\varphi _0}{h}\mathcal{I}[\frac{k_\alpha }{\epsilon ^{33}_P}]-A_{\mu \nu}\mathcal{I}[p_\alpha ^{\mu \nu }-\frac{k_\alpha }{\epsilon ^{33}_P}\langle \frac{1}{\epsilon_P^{33}} \rangle ^{-1}\langle q^{\mu \nu }\rangle ]
\\
+hB_{\mu \nu }\mathcal{I}[p_\alpha ^{\mu \nu}\zeta -\frac{k_\alpha }{\epsilon ^{33}_P}\langle \frac{1}{\epsilon_P^{33}} \rangle ^{-1}\langle q^{\mu \nu } \zeta \rangle ],
\end{multline}
with $\mathcal{I}[f]$ being defined as in the previous case. Finally, with $D$ from \eqref{5.21} being plugged in \eqref{5.20}, we obtain
\begin{multline}
\label{5.27}
\chi =-\frac{\varphi _0}{h}\int_{-1/2}^\zeta (1-\frac{1}{\epsilon_P^{33}(\xi )}\langle \frac{1}{\epsilon_P^{33}} \rangle ^{-1})d\xi +A_{\alpha \beta }\int_{-1/2}^\zeta (q^{\alpha \beta }(\xi )-\frac{\langle q^{\alpha \beta }\rangle }{\epsilon_P^{33}(\xi )}\langle \frac{1}{\epsilon_P^{33}} \rangle ^{-1})d\xi 
\\
-hB_{\alpha \beta } \int_{-1/2}^\zeta (q^{\alpha \beta }(\xi )\xi - \frac{\langle q^{\alpha \beta }\zeta \rangle}{\epsilon_P^{33}(\xi )}\langle \frac{1}{\epsilon_P^{33}} \rangle ^{-1})d\xi .
\end{multline}

\section{Two-dimensional theory}

\subsection{Unelectroded faces.}
In accordance with the variational-asymptotic method we take now the displacement field and the electric potential represented in \eqref{5.2}, where functions $y_\alpha $, $y$, and $\chi $ are given by \eqref{5.9}. We regard $u(x^\alpha ,t)$, $u_\alpha (x^\alpha ,t)$, and $\psi (x^\alpha ,t)$  as the unknown functions, with $A_{\alpha \beta }$ and $B_{\alpha \beta }$ describing the measures of extension and bending of the shell middle surface, respectively, and $F_\alpha $ the average electric field in the longitudinal directions. We substitute this displacement and electric potential into the action functional \eqref{3.4}. Since we construct the approximate theory admitting the error of order $h/R$, $\kappa $ in \eqref{3.4} may be replaced by 1. If we keep only the principal terms containing the unknown functions in the average Lagrangian and integrate over the thickness, then the average kinetic energy density becomes
\begin{equation}
\label{6.1}
\Theta (\dot{u}_\alpha ,\dot{u})=\frac{h}{2}\langle \rho (\zeta )\rangle (a^{\alpha \beta }\dot{u}_\alpha \dot{u}_\beta +\dot{u}^2).
\end{equation}
To compute the average electric enthalpy density we use the additive decomposition $W=W_\parallel +W_\perp $ that leads to
\begin{equation}
\label{6.2}
\Phi =h\langle W \rangle =h(\langle W_\parallel \rangle + \langle W_\perp \rangle ).
\end{equation}
On the fields \eqref{5.2} the average transverse electric enthalpy vanishes, while the principal terms of the average longitudinal electric enthalpy give
\begin{align}
\Phi &(A_{\alpha \beta },B_{\alpha \beta },F_\beta)=\frac{h}{2}\langle c^{\alpha \beta \gamma \delta }_N(\zeta )
(A_{\alpha \beta }-h\zeta B_{\alpha \beta })(A_{\gamma \delta }-h\zeta 
B_{\gamma \delta }) \notag
\\
&-2e^{\gamma \alpha \beta }_N(\zeta )
(A_{\alpha \beta }-h\zeta B_{\alpha \beta })F_\gamma -
\epsilon ^{\alpha \beta }_N(\zeta )F_\alpha F_\beta \rangle \notag
\\
&=\frac{h}{2}(\langle c^{\alpha \beta \gamma \delta }_N\rangle A_{\alpha \beta }
A_{\gamma \delta }-2h\langle c^{\alpha \beta \gamma \delta }_N\zeta \rangle A_{\alpha \beta }B_{\gamma \delta }+h^2\langle c^{\alpha \beta \gamma \delta }_N\zeta ^2\rangle 
B_{\alpha \beta }B_{\gamma \delta } \notag
\\
&-2\langle e^{\gamma \alpha \beta }_N\rangle A_{\alpha \beta }F_\gamma +2h\langle e^{\gamma \alpha \beta }_N\zeta \rangle B_{\alpha \beta }F_\gamma 
-\langle \epsilon ^{\alpha \beta }_N \rangle F_\alpha F_\beta ). \label{6.2a}
\end{align}
Thus, in contrast to the homogeneous piezoelectric shells, the FGP-shells exhibits in general the interaction between extension and bending due to the cross terms between $A_{\alpha \beta }$ and $B_{\alpha \beta }$ as will be seen below.

We formulate now the variational principle for the FGP-shell: the true average displacement field $\check{\mathbf{u}}(x^\alpha ,t)$ and electric potential $\check{\psi }(x^\alpha ,t)$ of the FGP-shell change in space and time in such a way that the 2-D average action functional
\begin{equation}
\label{6.3}
J[\mathbf{u}(x^\alpha ,t),\psi (x^\alpha ,t)]=\int_{t_0}^{t_1}\int_{\Omega }[\Theta (\dot{\mathbf{u}})-\Phi (A_{\alpha \beta },B_{\alpha \beta },F_\alpha )]\, da \, dt
\end{equation} 
becomes stationary among all continuously differentiable functions $\mathbf{u}(x^\alpha ,t)$ and $\psi (x^\alpha ,t)$ satisfying the initial and end conditions
\begin{displaymath}
\mathbf{u}(x^\alpha ,t_0)=\mathbf{u}_0(x^\alpha ), \quad \mathbf{u}(x^\alpha ,t_1)=\mathbf{u}_1(x^\alpha ),
\end{displaymath}
and the constraints 
\begin{equation}
\label{6.4}
\varphi =\varphi _{(i)}(t) \quad \text{on $c_e^{(i)}$}, \quad i=1,\ldots ,n.
\end{equation}
The standard calculus of variation shows that the stationarity condition $\delta J=0$ implies the following two-dimensional equations
\begin{align}
\langle \rho \rangle h\ddot{u}^\alpha &=T^{\alpha \beta }_{;
\beta }+b^\alpha _\lambda M^{\lambda \beta }_{;\beta } , \notag
\\
\langle \rho \rangle h\ddot{u}&=T^{\alpha \beta }b_{\alpha \beta }
-M^{\alpha \beta }_{;\alpha \beta },
\label{6.5}
\\
G^\alpha _{;\alpha }&=0,
\end{align}
where $T^{\alpha \beta }=N^{\alpha \beta }+b^\alpha _\lambda M^{\lambda \beta }$. These equations are subjected to the free-edge boundary conditions
\begin{align}
T^{\alpha \beta }\nu _{\beta }+b^\alpha _\gamma 
M^{\gamma \beta } \nu _\beta &=0, \notag \\ 
M^{\alpha \beta }_{;\alpha }\nu _\beta +\frac{\partial }{\partial s}(M^{\alpha \beta }
\tau _\alpha \nu _\beta )&=0,
\label{6.6} 
\\
M^{\alpha \beta }\nu _\alpha \nu _\beta &=0, \notag
\\
G^\alpha \nu _\alpha &=0 \quad \text{on $c_d$}, \notag
\end{align}
and constraints \eqref{6.4}, with $\nu _\alpha $ being the components of the unit surface vector normal to the curve $\partial \Omega$. For the clamped or simply supported edge, the natural boundary conditions in \eqref{6.6} must be replaced by the corresponding kinematical boundary conditions. The equations of motion \eqref{6.5} must be complemented by the constitutive equations
\begin{align}
N^{\alpha \beta }&= \frac{\partial \Phi }{\partial A_{\alpha \beta }}=h(\langle c^{\alpha \beta \gamma \delta }_N\rangle  A_{\gamma \delta }-h\langle c^{\alpha \beta \gamma \delta }_N\zeta \rangle B_{\gamma \delta }-\langle e^{\gamma \alpha \beta }_N\rangle F_\gamma ),
\notag 
\\
M^{\alpha \beta }&= \frac{\partial \Phi }{\partial B_{\alpha \beta }}=h^2(h\langle c^{\alpha \beta \gamma \delta }_N\zeta ^2\rangle B_{\gamma \delta }-\langle c^{\alpha \beta \gamma \delta }_N\zeta \rangle A_{\gamma \delta }+\langle e^{\gamma \alpha \beta }_N \zeta \rangle F_\gamma ), \label{6.7}
\\
G^\alpha &=-\frac{\partial \Phi }{\partial F_{\alpha }}=h(\langle \epsilon ^{\alpha \beta }_N \rangle F_\beta +\langle e^{\alpha \beta \gamma }_N\rangle A_{\beta \gamma }-h\langle e^{\alpha \beta \gamma }_N \zeta \rangle B_{\beta \gamma }).
\end{align}
From \eqref{6.7} we see that, in general, the extension measures cause the bending moments, and the bending measures cause the membrane stresses. This is the cross effect mentioned above. 

Within the framework of the first-order approximation of 2-D shell theories one can choose different measures of bending according to 
\begin{equation*}
\tilde{B}_{\alpha \beta}=B_{\alpha \beta }+C^{\gamma \delta }_{\alpha \beta }A_{\gamma \delta },
\end{equation*}
with $C^{\gamma \delta }_{\alpha \beta }$ being a linear function of $b_{\alpha \beta }$. For instance, the mostly used measures of bending $\rho _{\alpha \beta }$, proposed by \citet{koiter1960consistent} and \citet{Sanders59}, is related to $B_{\alpha \beta }$ by
\begin{equation*}
\rho _{\alpha \beta }=B_{\alpha \beta }-b^\gamma _{(\alpha }A_{\beta )\gamma }.
\end{equation*}
The corresponding energy densities differ from each other by small terms of the order $h/R$ compared with unity. Indeed, choosing for example Koiter-Sanders tensor $\rho _{\alpha \beta }$ 
instead of $B_{\alpha \beta }$ as the bending measures in the average electric enthalpy density \eqref{6.2a}, one can show that the average electric enthalpy densities differ from each other by cross terms of the type $\mu h^3bAB$. These terms are of the order $h/R$ compared with unity, since, due to the Cauchy-Schwarz inequality
\[
\mu h^3 bAB \le \mu h^2b(A^2+h^2B^2).
\]
Therefore, the average electric enthalpy densities are asymptotically equivalent within the first-order approximation. 

To complete the 2-D theory of FGP-shells we should also indicate the method of restoring the 3-D electroelastic state by means of the 2-D one. To do this, the strain tensor field $\boldsymbol{\varepsilon}(x^\alpha ,x,t)$ and the electric field $\mathbf{E}(x^\alpha ,x,t)$ should be found from \eqref{5.2} and \eqref{5.9}. Using the asymptotic formulas \eqref{5.6}, \eqref{5.7} in combination with \eqref{5.8}, we obtain
\begin{align}
\varepsilon _{\alpha \beta }&=A_{\alpha \beta }- h\zeta B_{
\alpha \beta },
\quad
2\varepsilon _{\alpha 3}= -p_{\alpha }^{\mu \nu }(\zeta )
(A_{\mu \nu }-h\zeta B_{\mu \nu })+p_\alpha ^\mu (\zeta )F_\mu , \notag
\\
\varepsilon _{33}&=-r^{\alpha \beta }(\zeta )(A_{\alpha \beta 
}-h\zeta B_{\alpha \beta })+r^\alpha (\zeta )F_\alpha  \label{6.7a}
\\
E_\alpha &=F_\alpha ,\quad E_3=-q^{\alpha \beta }(\zeta )(A_{\alpha \beta 
}-h\zeta B_{\alpha \beta })-q^\alpha (\zeta )F_\alpha \notag
\end{align}
The stress tensor field $\boldsymbol{\sigma }(x^\alpha ,x,t)$ and the electric induction field $\mathbf{D}(x^\alpha ,x,t)$ are then determined by the 3-D constitutive equations. While doing so, it is convenient to use the decomposition \eqref{4.1} for the electric enthalpy density. Within the first-order approximation we find 
\begin{gather}
\sigma^{\alpha \beta }=c^{\alpha \beta \gamma \delta }_N(\zeta )(A_{\gamma \delta }-hB_{\gamma \delta }\zeta )+\epsilon ^{\alpha \beta }_N(\zeta ) F_\beta ,
\quad
\sigma^{\alpha 3}=0,\quad \sigma^{33}=0,
\label{6.7b} 
\\
D^\alpha =e^{\alpha \beta \gamma }_N(\zeta )(A_{\beta \gamma }-hB_{\beta \gamma }\zeta )+\epsilon^{\alpha \beta }_N(\zeta )F_\beta ,\quad
D^3=0.
\notag
\end{gather}
All these formulas are accurate up to terms of the orders $h/R$ and $h/l$ of smallness. Note that
\begin{equation}
\label{6.7c}
\langle \sigma^{\alpha \beta }\rangle =\frac{N^{\alpha \beta }}{h}, \quad \langle \sigma^{\alpha \beta } \zeta \rangle =-\frac{M^{\alpha \beta }}{h^2}, \quad \langle D^\alpha \rangle =\frac{G^\alpha }{h}.
\end{equation}

\subsection{Electroded faces.}
In this case we take the displacement field and the electric potential from \eqref{5.11}, where functions $y_\alpha $, $y$, and $\chi $ are given by \eqref{5.25}, \eqref{5.26}, and \eqref{5.27}, respectively. We regard $u_\alpha (x^\alpha ,t)$ and $u(x^\alpha ,t)$ as the unknown functions, with $A_{\alpha \beta }$ and $B_{\alpha \beta }$ describing the measures of extension and bending of the shell middle surface, respectively. We substitute this displacement and electric potential into the action functional \eqref{3.4}. Again, $\kappa $ in \eqref{3.4} may be replaced by 1 within the first order approximation. If we keep only the principal terms containing the unknown functions in the average Lagrangian and integrate over the thickness, then the average kinetic energy density assumes exactly the same form \eqref{6.1}. To compute the average electric enthalpy density we use again the additive decomposition $W=W_\parallel +W_\perp $. As $E_\alpha $ is negligibly small on the fields \eqref{5.11}, we may neglect the last two terms of $W_\parallel $ in \eqref{4.2}$_1$. Then 
\begin{multline}
\label{6.10}
\langle W_\parallel \rangle =\frac{h}{2}\langle c^{\alpha \beta \gamma \delta }_N(\zeta )
(A_{\alpha \beta }-h\zeta B_{\alpha \beta })(A_{\gamma \delta }-h\zeta 
B_{\gamma \delta })\rangle 
\\
=\frac{h}{2}(\langle c^{\alpha \beta \gamma \delta }_N\rangle A_{\alpha \beta }
A_{\gamma \delta }-2h\langle c^{\alpha \beta \gamma \delta }_N\zeta \rangle A_{\alpha \beta }B_{\gamma \delta }+h^2\langle c^{\alpha \beta \gamma \delta }_N\zeta ^2\rangle 
B_{\alpha \beta }B_{\gamma \delta }) .
\end{multline}
For the average transverse electric enthalpy density we substitute formulas $\gamma =f(\zeta )F$, $\gamma _\alpha =k_\alpha (\zeta )F$ into \eqref{5.14}. This yields
\begin{equation}
\label{6.11}
\langle W_\perp \rangle =-\frac{h}{2}\langle \epsilon ^{33}_P(\zeta )F^2 \rangle =-\frac{h}{2}\langle \frac{D^2}{\epsilon ^{33}_P(\zeta )} \rangle .
\end{equation}
With $D$ from \eqref{5.21}, we arrive at
\begin{equation}
\label{6.12}
\langle W_\perp \rangle =-\frac{h}{2}\langle \frac{1}{\epsilon^{33}_P} \rangle ^{-1} (-\frac{\varphi _0(t)}{h}+A_{\alpha \beta }\langle q^{\alpha \beta }\rangle -hB_{\alpha \beta } \langle q^{\alpha \beta }\zeta \rangle )^2.
\end{equation}
Combining the average longitudinal and transverse electric enthalpy densities together, we obtain 
\begin{multline}
\label{6.13}
\Phi (A_{\alpha \beta },B_{\alpha \beta })=\frac{h}{2}(\langle c^{\alpha \beta \gamma \delta }_N\rangle A_{\alpha \beta }
A_{\gamma \delta }-2h\langle c^{\alpha \beta \gamma \delta }_N\zeta \rangle A_{\alpha \beta }B_{\gamma \delta }+h^2\langle c^{\alpha \beta \gamma \delta }_N\zeta ^2\rangle 
B_{\alpha \beta }B_{\gamma \delta } )
\\
-\frac{h}{2}\langle \frac{1}{\epsilon^{33}_P} \rangle ^{-1} (-\frac{\varphi _0(t)}{h}+A_{\alpha \beta }\langle q^{\alpha \beta }\rangle -hB_{\alpha \beta } \langle q^{\alpha \beta }\zeta \rangle )^2.
\end{multline}
Looking at \eqref{6.13} we recognize again the interaction and cross effects between extension, bending, and external electric field for the FGP-shells.

The variational principle for the FGP-shell with electroded faces states that the true average displacement field $\check{\mathbf{u}}(x^\alpha ,t)$ of the FGP-shell change in space and time in such a way that the 2-D average action functional
\begin{equation}
\label{6.14}
J[\mathbf{u}(x^\alpha ,t)]=\int_{t_0}^{t_1}\int_{\Omega }[\Theta (\dot{\mathbf{u}})-\Phi (A_{\alpha \beta },B_{\alpha \beta })]\, da \, dt
\end{equation} 
becomes stationary among all continuously differentiable functions $\mathbf{u}(x^\alpha ,t)$  satisfying the initial and end conditions
\begin{displaymath}
\mathbf{u}(x^\alpha ,t_0)=\mathbf{u}_0(x^\alpha ), \quad \mathbf{u}(x^\alpha ,t_1)=\mathbf{u}_1(x^\alpha ).
\end{displaymath}
Thus, this variational principle yields the same equations of motion and boundary conditions as in the theory of inhomogeneous elastic shells (equations \eqref{6.5}$_{1,2}$ and boundary conditions \eqref{6.6}$_{1,2,3}$). The only changes concern the constitutive equations obtained from the average electric enthalpy \eqref{6.13}. For the FGP-shell we have
\begin{align}
N^{\alpha \beta }= \frac{\partial \Phi }{\partial A_{\alpha \beta }}&=h[ \langle c^{\alpha \beta \gamma \delta }_N\rangle  A_{\gamma \delta }-h\langle c^{\alpha \beta \gamma \delta }_N\zeta \rangle B_{\gamma \delta } \notag
\\
&-\langle q^{\alpha \beta }\rangle  \langle \frac{1}{\epsilon^{33}_P} \rangle ^{-1} (-\frac{\varphi _0(t)}{h}+A_{\gamma \delta }\langle q^{\gamma \delta }\rangle -hB_{\gamma \delta } \langle q^{\gamma \delta }\zeta \rangle )], \label{6.14a}
\\
M^{\alpha \beta }= \frac{\partial \Phi }{\partial B_{\alpha \beta }}&=h^2[h\langle c^{\alpha \beta \gamma \delta }_N\zeta ^2\rangle B_{\gamma \delta }-\langle c^{\alpha \beta \gamma \delta }_N\zeta \rangle A_{\gamma \delta } \notag
\\
&+\langle q^{\alpha \beta }\zeta \rangle  \langle \frac{1}{\epsilon^{33}_P} \rangle ^{-1} (-\frac{\varphi _0(t)}{h}+A_{\gamma \delta }\langle q^{\gamma \delta }\rangle -hB_{\gamma \delta } \langle q^{\gamma \delta }\zeta \rangle )]. \label{6.14b}
\end{align}
Similar to the previous case, one can use different bending measure in the average electric enthalpy \eqref{6.13} leading to the asymptotically equivalent governing equations. However, as shown by \citet{le1979on}, only some of those bending measures ensure the static-geometric analogy for 2-D shell theory.

The reconstruction of the 3-D electroelastic state by means of the 2-D one can be done in the similar manner. First, the strain tensor field $\boldsymbol{\varepsilon}(x^\alpha ,x,t)$ and the electric field $\mathbf{E}(x^\alpha ,x,t)$ should be restored in accordance with \eqref{5.13}. Using \eqref{5.23}, we find that
\begin{multline}
\label{6.15}
\varepsilon _{33}=y_{|\zeta }=-\frac{f(\zeta )}{\epsilon ^{33}_P(\zeta )}\langle \frac{1}{\epsilon_P^{33}} \rangle ^{-1}\frac{\varphi _0}{h}-A_{\alpha \beta }(r^{\alpha \beta}(\zeta )-\frac{f(\zeta )}{\epsilon ^{33}_P(\zeta )}\langle \frac{1}{\epsilon_P^{33}} \rangle ^{-1}\langle q^{\alpha \beta }\rangle )
\\
+hB_{\alpha \beta }(r^{\alpha \beta}(\zeta )\zeta -\frac{f(\zeta )}{\epsilon ^{33}_P(\zeta )}\langle \frac{1}{\epsilon_P^{33}} \rangle ^{-1}\langle q^{\alpha \beta } \zeta \rangle ),
\end{multline}
Similarly, from \eqref{5.24} follows
\begin{multline}
\label{6.16}
2\varepsilon _{\alpha 3}=y_{\alpha |\zeta }=-\frac{k_\alpha (\zeta )}{\epsilon ^{33}_P(\zeta )}\langle \frac{1}{\epsilon_P^{33}} \rangle ^{-1}\frac{\varphi _0}{h}-A_{\mu \nu}(p_\alpha ^{\mu \nu }(\zeta )-\frac{k_\alpha (\zeta )}{\epsilon ^{33}_P(\zeta )}\langle \frac{1}{\epsilon_P^{33}} \rangle ^{-1}\langle q^{\mu \nu }\rangle )
\\
+hB_{\mu \nu }(p_\alpha ^{\mu \nu}(\zeta )\zeta -\frac{k_\alpha (\zeta )}{\epsilon ^{33}_P(\zeta )}\langle \frac{1}{\epsilon_P^{33}} \rangle ^{-1}\langle q^{\mu \nu } \zeta \rangle ).
\end{multline}
For the non-zero component $E_3$ of the electric field we have
\begin{multline}
\label{6.16a}
E_3=-\frac{1}{\epsilon ^{33}_P(\zeta )}\langle \frac{1}{\epsilon_P^{33}} \rangle ^{-1}\frac{\varphi _0}{h}-A_{\mu \nu}(q^{\alpha \beta }(\zeta )-\frac{1}{\epsilon ^{33}_P(\zeta )}\langle \frac{1}{\epsilon_P^{33}} \rangle ^{-1}\langle q^{\alpha \beta }\rangle )
\\
+hB_{\alpha \beta }(q^{\alpha \beta }(\zeta )\zeta -\frac{1}{\epsilon ^{33}_P(\zeta )}\langle \frac{1}{\epsilon_P^{33}} \rangle ^{-1}\langle q^{\alpha \beta } \zeta \rangle ).
\end{multline}
The stress tensor field $\boldsymbol{\sigma }(x^\alpha ,x,t)$ and the electric induction field $\mathbf{D}(x^\alpha ,x,t)$ are then determined by the 3-D constitutive equations. While doing so, it is convenient to use the decomposition \eqref{4.1} for the electric enthalpy density. Within the first-order approximation we find the components of the stress tensor field
\begin{align}
\sigma^{\alpha \beta }&=c^{\alpha \beta \gamma \delta }_N(\zeta )(A_{\gamma \delta }-hB_{\gamma \delta }\zeta )
\\
&-q^{\alpha \beta }(\zeta ) \langle \frac{1}{\epsilon^{33}_P} \rangle ^{-1} (-\frac{\varphi _0(t)}{h}+A_{\gamma \delta }\langle q^{\gamma \delta }\rangle -hB_{\gamma \delta } \langle q^{\gamma \delta }\zeta \rangle ) \label{6.17} 
\\ 
\sigma^{\alpha 3}&=0,\quad \sigma^{33}=0.
\end{align}
For the components of the electric induction field we have 
\begin{equation}\label{6.18}
D^\alpha =0 ,\quad
D^3=\langle \frac{1}{\epsilon_P^{33}(\zeta )} \rangle ^{-1}(-\frac{\varphi _0}{h}+A_{\alpha \beta }\langle q^{\alpha \beta }(\zeta )\rangle -hB_{\alpha \beta } \langle q^{\alpha \beta }(\zeta )\zeta \rangle ).
\end{equation}
Note that
\begin{equation}
\label{6.19}
\langle \sigma^{\alpha \beta }\rangle =\frac{N^{\alpha \beta }}{h}, \quad \langle \sigma^{\alpha \beta } \zeta \rangle =-\frac{M^{\alpha \beta }}{h^2}.
\end{equation}
Again, these formulas are accurate up to terms of the orders $h/R$ and $h/l$ of smallness. It is easy to check that all equations and formulas in Sections 5 and 6 reduce to those of the homogeneous piezoelectric shells obtained by \citet{Le86a} and of the piezoelectric sandwich shells obtained by \citet{le2016asymptotically}.

\section{Error estimation of the constructed 2-D theory}
In this Section we shall use the identity found by \citet{le2016asymptotically} for inhomogeneous piezoelectric bodies to give an error estimate of the functionally graded piezoelectric shell theory constructed in the previous Section in the special case of statics. 

We consider an inhomogeneous piezoelectric body occupying the three-dimensional domain $\mathcal{V}$ in its undeformed state that stays in equilibrium under a fixed voltage. Concerning the boundary conditions for the mechanical quantities we assume that the boundary $\partial \mathcal{V}$ is decomposed into two subboundaries $\mathcal{S}_k$ and $\mathcal{S}_s$. On the part $\mathcal{S}_k$ the displacements vanish (clamped boundary)
\begin{equation}
\mathbf{w}=0 \quad \text{on $\mathcal{S}_k$}.
\label{7.1}
\end{equation}
On the remaining part $\mathcal{S}_s$ the traction-free boundary condition is assumed
\begin{equation}
\boldsymbol{\sigma }\cdot \mathbf{n} = \mathbf{0} \quad \text{on
$\mathcal{S}_s$}.
\label{7.2}
\end{equation}
Concerning the boundary conditions for the electric potential we assume that the boundary $\partial \mathcal{V}$ consists of $n+1$ subboundaries $\mathcal{S}_e^{(1)}, \ldots , \mathcal{S}_e^{(n)}$ and $\mathcal{S}_d$. The subboundaries $\mathcal{S}_e^{(1)},\ldots , \mathcal{S}_e^{(n)}$ are covered by electrodes with negligible thickness. On these electrodes the electric potential is prescribed
\begin{equation}
\varphi =\varphi _{(i)} \quad \text{on
$\mathcal{S}_e^{(i)},i=1,\ldots ,n$}.
\label{7.3}
\end{equation}
On the uncoated portion $\mathcal{S}_d$ of the boundary we require that the surface charge vanishes
\begin{equation}
\mathbf{D}\cdot \mathbf{n} = 0 \quad \text{on $\mathcal{S}_d$}.
\label{7.4}
\end{equation}

We introduce the linear vector space of electroelastic states that consists of elements of the 
form $\boldsymbol{\Xi} =(\boldsymbol{\sigma },\mathbf{E})$, where $\boldsymbol{\sigma }$ is the stress field and $\mathbf{E}$ is the electric field; both fields are defined in the three-dimensional domain $\mathcal{V}$ occupied by the piezoelectric body. In this space we introduce the following energetic norm
\begin{equation}
\parallel \boldsymbol{\Xi} \parallel ^2_{L_2}=C_2[\boldsymbol{\Xi} ]=\int_{\mathcal{V}}
G(\mathbf{x},\boldsymbol{\sigma },\mathbf{E})\, dv,
\label{7.5}
\end{equation}
where function $G(\mathbf{x},\boldsymbol{\sigma },\mathbf{E})$ is the density of the complementary energy (or Gibbs function) of the inhomogeneous piezoelectric body. In the index notation $G(\mathbf{x},\boldsymbol{\sigma },\mathbf{E})$ reads
\begin{equation*}
G(\mathbf{x},\boldsymbol{\sigma },\mathbf{E})=\frac{1}{2}s^E_{abcd}(\mathbf{x})\sigma^{ab}\sigma^{cd}+d_{cab}(\mathbf{x})\sigma^{ab}E^c+\frac{1}{2}\epsilon ^T_{ab}(\mathbf{x})E^aE^b.
\end{equation*}
Since the complementary energy density $G(\mathbf{x},\boldsymbol{\sigma },\mathbf{E})$ is positive definite, the definition \eqref{7.5} is meaningful. 

We call ``kinematically admissible'' those electroelastic states $\check{\boldsymbol{\Xi} }$ for which the compatible strain field $\check{\boldsymbol{\varepsilon }}$ and the electric induction field $\check{\mathbf{D}}$ exist such that
\begin{gather*}
\check{\boldsymbol{\varepsilon }}=\frac{1}{2}(\nabla \check{\mathbf{w}}+(\nabla 
\check{\mathbf{w}})^T), \quad \check{\mathbf{w}}=0 \quad 
\text{on $\mathcal{S}_k$},
\\
\text{div}\check{\mathbf{D}}=0,\quad
\check{\mathbf{D}}\cdot \mathbf{n}=0 \quad \text{on $\mathcal{S}_d$},
\end{gather*}
while $\check{\boldsymbol{\sigma }}$ and $\check{\mathbf{E}}$ are expressed in terms of $\check{\boldsymbol{\varepsilon }}$ and $\check{\mathbf{D}}$ by the constitutive equations equivalent to \eqref{2.6}. We call those electroelastic states $\hat{\boldsymbol{\Xi} }$ ``statically admissible'', when
\begin{gather*}
\text{div}\hat{\boldsymbol{\sigma }}=0,\quad \hat{\boldsymbol{\sigma }}\cdot \mathbf{n}=0 \quad
\text{on $\mathcal{S}_s$},
\\
\hat{\mathbf{E}}=-\nabla \hat{\varphi }, \quad \hat{\varphi 
}=\varphi _{(i)} \quad \text{on $\mathcal{S}_e^{(i)},i=1,\ldots 
,n$}.
\end{gather*}

Let $\tilde{\boldsymbol{\Xi} }=(\tilde{\boldsymbol{\sigma }},\tilde{\mathbf{E}})$ be the true 
electroelastic state that is realized in an inhomogeneous piezoelectric body staying in equilibrium under the given (time-independent) values of the electric potential $\varphi _{(i)}$ on the electrodes. Then the following identity
\begin{equation}
C_2[\tilde{\boldsymbol{\Xi}}-\frac{1}{2}(\check{\boldsymbol{\Xi} }+\hat{\boldsymbol{\Xi}})]=
C_2[\frac{1}{2}(\check{\boldsymbol{\Xi}}-\hat{\boldsymbol{\Xi}})]
\label{7.7}
\end{equation}
turns out to be valid for arbitrary kinematically and statically admissible fields $\check{\boldsymbol{\Xi} }$ and $\hat{\boldsymbol{\Xi} }$. This identity generalizes the well-known Prager-Synge identity \citep{prager1947approximations} to the statics of inhomogeneous piezoelectric bodies. It implies that $\frac{1}{2}(\check{\boldsymbol{\Xi} }+\hat{\boldsymbol{\Xi} })$ may be regarded as an ``approximation'' to the true solution in the energetic norm, provided the complementary energy associated with the difference $\frac{1}{2}(\check{\boldsymbol{\Xi} }-\hat{\boldsymbol{\Xi} })$ is ``small''. In this case we may also consider each of the fields $\check{\boldsymbol{\Xi} }$ or $\hat{\boldsymbol{\Xi} }$ as an ``approximation'', in view of the inequalities
\begin{gather*}
C_2[\tilde{\boldsymbol{\Xi} }-\check{\boldsymbol{\Xi} })]\le C_2[\check{\boldsymbol{\Xi} }-\hat{\boldsymbol{\Xi} }],
\\
C_2[\tilde{\boldsymbol{\Xi} }-\hat{\boldsymbol{\Xi} })]\le C_2[\check{\boldsymbol{\Xi} }-\hat{\boldsymbol{\Xi} }],
\end{gather*}
which follow easily from \eqref{7.7}. For homogeneous piezoelectric bodies the identity \eqref{7.7} has been proven  in \citep{Le86a}, while for inhomogeneous piezoelectric bodies it has been established in our recent paper \citep{le2016asymptotically}.

Based on \eqref{7.7} the following error estimate can be established.\footnote{This error estimation generalizes the results obtained first by \citet{koiter1970mathematical} for the elastic shells, by \citet{Le86a} for the homogeneous piezoelectric shells, and by \citet{le2016asymptotically} for the piezoelectric sandwich shells.}

{\it Theorem.} The electroelastic state determined by the 2-D static theory of functionally graded piezoelectric shells differs in the norm $L_2$ from the exact electroelastic state determined by the 3-D theory of piezoelectricity by a quantity of the order $h/R+h/l$ as compared with unity.

To prove this theorem it is enough to find out the kinematically and statically admissible 3-D fields of electroelastic states that differ from the electroelastic state determined by the 2-D theory by a quantity of the order $h/R+h/l$ as compared with unity. Below we shall construct these 
fields for the two cases of unelectroded and electroded faces separately.

\subsection{Unelectroded faces.}

{\it Construction of kinematically admissible field.} We specify the kinematically admissible 
displacement field in the form
\begin{align*}
\check{w}_\alpha (x^\alpha ,x)&=u_\alpha (x^\alpha )-x \varphi _{\alpha } 
(x^\alpha ) +hy_\alpha (x^\alpha ,x),
\\
\check{w}(x^\alpha ,x)&=u(x^\alpha )+hy (x^\alpha ,x),
\end{align*}
where $\varphi _\alpha =u_{,\alpha }+b_\alpha ^\mu u_\mu $, while $y(x^\alpha ,x)$ and $y_\alpha (x^\alpha ,x)$ are given by \eqref{5.9}. Here and below, all quantities without the superscripts $\hat{\null }$ and $\check{\null }$ refer to the solution of the equilibrium equations of FGP-shells obtained by the constructed two-dimensional theory. The components of the strain tensor are calculated according to the exact 3-D kinematic formulas \eqref{3.5}. Assume that the 2-D electroelastic state is characterized by the strain amplitude $\varepsilon =\varepsilon _A+\varepsilon _B$, and the quantity $f_F$ defined in Section 5 is expressed through $\varepsilon $ by $f_F=c\varepsilon $, with $c$ a constant. The asymptotic analysis similar to that given in Section 5 shows that
\begin{equation*}
\check{\boldsymbol{\varepsilon }}(x^\alpha ,x)=\boldsymbol{\varepsilon }(x^\alpha ,x)+O(h/R,h/l)\varepsilon ,
\end{equation*}
with $\boldsymbol{\varepsilon }(x^\alpha ,x)$ from \eqref{6.7a}. We choose the components $\check{D}^\alpha $ of the electric induction field to be equal to $D^\alpha /\kappa $, with $D^\alpha $ from \eqref{6.7b}$_3$. The components $\check{D}^3$ must be found by solving the 3-D equation of electrostatics
\begin{equation}
(\check{D}^\alpha \kappa )_{;\alpha }+(\check{D}^3\kappa 
)_{,x}=0,
\label{7.8}
\end{equation}
subject to the boundary conditions $\check{D}^3=0$ at $x=\pm h/2$. Due to the above choice for $\check{D}^\alpha $ equation \eqref{7.8} yields an unique solution
\begin{eqnarray*}
\check{D}^3=\frac{1}{\kappa }\int_{-h/2}^x(\check{D}^\alpha (\xi 
)\kappa (\xi ))_{;\alpha }\, d\xi =O(h/R,h/l)\epsilon
\end{eqnarray*}
that satisfies the boundary conditions at $x=\pm h/2$ (the condition $\check{D}^3=0$ at $x= 
h/2$ is fulfilled because of the 2-D equation of electrostatics $G^\alpha _{;\alpha }=0$). Note that the constructed field $\check{\bf D}(x^\alpha ,x)$ does not satisfy the exact boundary condition $\check{D}^\alpha \kappa \nu _\alpha =0$, posed at the portion $c_d\times [-h/2,h/2]$ of the edge, but satisfies it only on ``average'', i.e.
\[
\langle \check{D}^\alpha \kappa \rangle _x\nu _\alpha =
\frac{G^\alpha }{h}\nu _\alpha =0.
\]
For simplicity of the proof we further assume that the 3-D boundary conditions at the edge of the shell agree with the inner expansion of the electroelastic state (the so-called regular boundary conditions). Then the electric induction field $\check{\bf D}(x^\alpha ,x)$ constructed above is kinematically admissible. For irregular boundary conditions we have to take into account an additional electric induction field that differs substantially from zero only in a thin boundary layer at the shell edge. Since the energy of this boundary layer is of the order $h/l$ compared with that of the inner domain, one can easily generalize the proof of the theorem to this case.

Knowing $(\check{\boldsymbol{\varepsilon }},\check{\mathbf{D}})$, we find $\check{\boldsymbol{\Xi} }=(\check{\boldsymbol{\sigma }},\check{\mathbf{E}})$ from the constitutive equations equivalent to \eqref{2.6}. Because $(\check{\boldsymbol{\varepsilon }},\check{\mathbf{D}})=(\boldsymbol{\varepsilon },\mathbf{D})+O(h/R,h/l)\varepsilon $, it is easily seen that $(\check{\boldsymbol{\sigma }},\check{\mathbf{E}})=(\boldsymbol{\sigma },\mathbf{E})+O(h/R,h/l)\varepsilon$.

{\it Construction of statically admissible field.} We write down the exact 3-D equilibrium equations for a shell in the form (cf. \citep{le1999vibrations})
\begin{equation}
\begin{split}
\hat{\tau }^{\alpha \beta }_{;\beta }
+(\mu ^\alpha _\beta \hat{\tau }^\beta )_{,x}-\hat{\tau }^\beta b^\alpha 
_\beta =0,
\\
\hat{\tau }^\beta _{;\beta }+\hat{\tau }^{\alpha 
\beta }b_{\alpha \beta }+\hat{\tau }_{,x}=0,
\end{split}
\label{7.9}
\end{equation}
where
$$\hat{\tau }^{\alpha \beta }=\mu ^\alpha _\lambda \hat{\sigma }
^{\lambda \beta }\kappa ,\quad \hat{\tau }^\alpha =\hat{\sigma }^{\alpha 3}
\kappa ,\quad \hat{\tau }=\hat{\sigma }^{33}\kappa .$$
Note that $\hat{\tau }^{\alpha \beta }$ is unsymmetric. To find the statically admissible stress field $\hat{\boldsymbol{\sigma }}(x^\alpha ,x)$ satisfying \eqref{7.9} and the traction-free boundary conditions
\begin{equation}
\mu ^\alpha _\beta \hat{\sigma }^{\beta 3}\kappa =0,\quad
\hat{\sigma }^{33}\kappa =0 \quad \text{at $x=\pm h/2$,}
\label{7.10}
\end{equation}
we proceed as follows. We decompose the stress tensor field $\sigma ^{\alpha \beta }(x^\alpha ,x)$ from \eqref{6.7b}$_1$ into the sum
\begin{displaymath}
\sigma ^{\alpha \beta }(x^\alpha ,x)=\sigma ^{\alpha \beta }_1(x^\alpha ,x)+\sigma ^{\alpha \beta }_2(x^\alpha ,x),
\end{displaymath}
where $\sigma ^{\alpha \beta }_1(x^\alpha ,x)$ is even in $x$, while $\sigma ^{\alpha \beta }_2(x^\alpha ,x)$ is odd in $x$. Let $\lambda ^\alpha _\beta (x)$ be the inverse matrix to $\mu ^\alpha _\beta (x)=\delta ^\alpha _\beta -b^\alpha _\beta x$ such that $\mu ^\alpha _\beta (x)\lambda ^\beta _\gamma (x)=\delta ^\alpha _\gamma $. It is easy to check that
\begin{displaymath}
\lambda ^\alpha _\beta (x)=\frac{1}{\kappa }[(1-2Hx)\delta ^\alpha _\beta +xb^\alpha _\beta ] .
\end{displaymath}
We choose $\hat{\sigma }^{\alpha \beta}(x^\alpha ,x)$ as follows
\begin{equation*}
\hat{\sigma}^{\alpha \beta }(x^\alpha ,x)=\lambda ^{(\alpha }_\gamma \sigma ^{\gamma \beta )}_1/\kappa +\sigma ^{\alpha \beta }_2/\kappa .
\end{equation*}
It is easily seen that
\begin{equation}
\langle \hat{\tau }^{\alpha \beta }\rangle _x=T^{\alpha \beta },
\quad \langle \hat{\tau }^{\alpha \beta }x\rangle _x=-M^{\alpha 
\beta },
\label{7.12}
\end{equation}
where $\langle . \rangle _x=\int . dx/h$. Besides, due to the property of $\lambda ^\alpha _\beta (x)$ and $\kappa (x)$ we have
\begin{equation*}
\hat{\sigma }^{\alpha \beta }(x^\alpha ,x)=\sigma ^{\alpha \beta }(x^\alpha ,x)+O(h/R,h/l)\varepsilon .
\end{equation*}
Solving \eqref{7.9},\eqref{7.10} with the given $\hat{\tau }^{\alpha \beta }$, we can find $\hat{\tau }^\alpha $ and $\hat{\tau }$ and then $\hat{\sigma}^{\alpha 3}$ and $\hat{\sigma }^{33}$. It turns out that \eqref{7.12} are the sufficient conditions for the existence of $\hat{\tau }^\alpha $ and $\hat{\tau }$. Indeed, integrating \eqref{7.9}$_1$ and \eqref{7.9}$_2$ multiplied by $x$ over $x\in [-h/2,h/2]$, we obtain
\begin{equation}
\begin{split}
T^{\alpha \beta }_{;\beta }-b^\alpha _\beta N^\beta +(\mu ^\alpha _\beta 
\hat{\tau }^\beta )|_{-h/2}^{h/2}=0,
\\
N^\alpha _{;\alpha }+b_{\alpha \beta }T^{\alpha \beta }+\hat{\tau }|_
{-h/2}^{h/2}=0,
\\
-M^{\alpha \beta }_{;\beta }-N^\alpha +(x\mu ^\alpha _\beta 
\hat{\tau }^\beta )|_{-h/2}^{h/2}=0,
\end{split}
\label{7.13}
\end{equation}
where $N^\alpha =\langle \hat{\tau }^\alpha \rangle _x$. From the first and the last equations of \eqref{7.13} it follows that $(\mu ^\alpha _\beta \hat{\tau }^\beta )|^{h/2}_{-h/2}=0$, since
$T^{\alpha \beta }_{;\beta }+b^\alpha _\lambda M^{\lambda \beta }_{;\beta }=0$ according to the 2-D equations of equilibrium. From the second equation of \eqref{7.13} we also obtain 
$\hat{\tau }|^{h/2}_{-h/2}=0$. Thus, if the boundary conditions \eqref{7.10} are satisfied at $x=-h/2$, then after the integration they will also be satisfied at $x=h/2$. Not showing the cumbersome solution of \eqref{7.9}, we note only that $\hat{\sigma}^{\alpha 3}, \hat{\sigma}^{33}\sim O(h/R,h/l)\varepsilon$. Thus, $\hat{\boldsymbol{\sigma }}(x^\alpha ,x)=\boldsymbol{\sigma }(x^\alpha ,x)
+O(h/R,h/l)\varepsilon $.

Concerning the statically admissible electric field $\hat{\mathbf{E}}(x^\alpha ,x)$ we specify its potential by
\begin{equation*}
\hat{\varphi }(x^\alpha ,x)=\psi (x^\alpha )+ h\chi (x^\alpha ,x),
\end{equation*}
with $\chi (x^\alpha ,x)$ from \eqref{5.9}. Then
\begin{equation*}
\hat{\mathbf{E}}(x^\alpha ,x)= \mathbf{E}(x^\alpha ,x)+O(h/l)\varepsilon ,
\end{equation*}
with $\mathbf{E}(x^\alpha ,x)$ from \eqref{6.7a}$_3$. Note that the statically admissible field $(\hat{\boldsymbol{\sigma }},\hat{\mathbf{E}})$ constructed above satisfies only the regular boundary conditions at the shell edge, exactly as in the previous case.

\subsection{Electroded faces.}

{\it Construction of kinematically admissible field.} The displacement field $\hat{\mathbf{w}}(x^\alpha ,x)$ is taken as in \eqref{5.11}, with $y(x^\alpha ,x)$ and $y_\alpha (x^\alpha ,x)$ from \eqref{5.25} and \eqref{5.26}, respectively. The components of the strain tensor field are calculated according to \eqref{3.5}. Assume that the 2-D electroelastic state is characterized by the strain amplitude $\varepsilon =\varepsilon _A+\varepsilon _B$. The asymptotic analysis similar to that given in Section 5 shows that
\begin{equation*}
\check{\boldsymbol{\varepsilon }}(x^\alpha ,x)=\boldsymbol{\varepsilon }(x^\alpha ,x)+O(h/R,h/l)\varepsilon ,
\end{equation*}
with $\boldsymbol{\varepsilon }(x^\alpha ,x)$ from \eqref{6.15} and \eqref{6.16}. We choose the components $\check{D}^\alpha $ of the electric induction field to be zero, while
\begin{equation}
\check{D}^3(x^\alpha ,x)=D^3(x^\alpha )/\kappa (x),
\end{equation}
with $D^3(x^\alpha )$ from \eqref{6.18}. It is easy to see that $\check{\mathbf{D}}(x^\alpha ,x)$ satisfies the exact 3-D equation of electrostatics
\begin{equation}
(\check{D}^\alpha \kappa )_{;\alpha }+(\check{D}^3\kappa 
)_{,x}=0.
\label{7.14}
\end{equation}
and, due to the property of $\kappa $,
\begin{equation*}
\check{\mathbf{D}}(x^\alpha ,x)=\mathbf{D}(x^\alpha )+O(h/R,h/l)\epsilon .
\end{equation*}
Knowing $(\check{\boldsymbol{\varepsilon }},\check{\mathbf{D}})$, we find $\check{\boldsymbol{\Xi} }=(\check{\boldsymbol{\sigma }},\check{\mathbf{E}})$ from the constitutive equations equivalent to \eqref{2.6}. Because $(\check{\boldsymbol{\varepsilon }},\check{\mathbf{D}})=(\boldsymbol{\varepsilon },\mathbf{D})+O(h/R,h/l)\varepsilon $, it is easily seen that $(\check{\boldsymbol{\sigma }},\check{\mathbf{E}})=(\boldsymbol{\sigma },\mathbf{E})+O(h/R,h/l)\varepsilon$.

{\it Construction of statically admissible field.} The statically admissible electric field $\hat{\mathbf{E}}(x^\alpha ,x)$ is derivable from its potential 
\begin{equation*}
\hat{\varphi }(x^\alpha ,x)=\frac{\varphi _0}{h}x+h\chi (x^\alpha ,x),
\end{equation*}
with $\chi (x^\alpha ,x)$ from \eqref{5.20}. Then
\begin{equation*}
\hat{E}_3(x^\alpha ,x)= E_3(x^\alpha ,x),
\quad
\hat{E}_\alpha (x^\alpha ,x)=O(h/l)\varepsilon .
\end{equation*}
The tensor field $\hat{\boldsymbol{\sigma }}(x^\alpha ,x)$ is constructed in the same way as
in case (i), and the asymptotic formula $\hat{\boldsymbol{\sigma }}(x^\alpha ,x)=\boldsymbol{\sigma }(x^\alpha ,x)+O(h/l)\varepsilon $ can be proved in a similar manner.

From this construction we see that $\check{\boldsymbol{\Xi} }$ and $\hat{\boldsymbol{\Xi} }$ differ from those found by the 2-D theory by a quantity of the order $h/R$ and $h/l$ as compared with unity. We have thus established the asymptotic accuracy of the electroelastic state determined by the 2-D theory in the energetic norm \eqref{7.5}. Note, however, that this error estimation does not alway guarantee the accuracy of the displacements. The reason is that in some problems $A_{\alpha \beta }\gg B_{\alpha \beta}$, so the bending measures belong to the category of small correction terms \citep{berdichevsky1992effect}. The first order approximate 2-D shell theory cannot determine the correction terms accurately, and consequently, the accuracy in determining displacements could be achieved only in the refined shell theories.

\section{Axisymmetric vibration of cylindrical FGP-shells}

\begin{figure}[htb]
    \begin{center}
    \includegraphics[height=5cm]{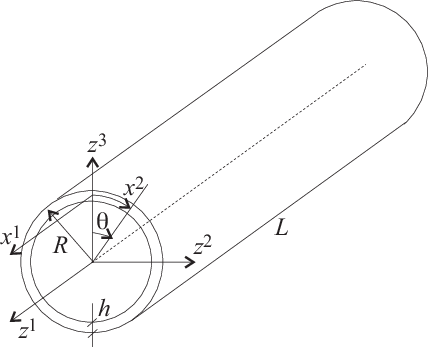}
    \end{center}
    \caption{A half of circular cylindrical shell.}
    \label{fig:3}
\end{figure}

In this Section we illustrate the application of the theory to the problem of forced axisymmetric vibration of a functionally graded piezoceramic cylindrical shell with thickness polarization fully covered by the electrodes. We refer the middle surface of the circular cylindrical shell having the radius $R$ and the length $2L$ to the curvilinear coordinates $\{x^1, x^2\}$, where $x^1\in (-L,L)$ is directed along the cylinder axis, while $x^2=R\theta $ is the circumferential coordinate (see Fig.~\ref{fig:3}). When this FG piezoceramic shell is subjected to an oscillating voltage, the electric field occurs causing its axisymmetric vibration for which $u_2=0$, while $u_1$ and $u$ do not depend on $x^2$. Under this condition the measures of extension and bending are given by
\begin{gather*}
A_{11}=u_{1,1},\quad A_{12}=A_{21}=0,
\quad A_{22}=\frac{u}{R},
\\
B_{11}=u_{,11},\quad B_{12}=B_{21}=0,
\quad B_{22}=0.
\end{gather*}
Since the piezoceramic material with the thickness polarization possesses the transverse isotropy, all the 2-D tensors of electroelastic moduli of odd rank vanish, in particular $e^{\alpha \beta \gamma }_N=0$. Besides,
\begin{equation*}
c^{\alpha \beta \gamma \delta }_N=c^N_1a^{\alpha \beta }a^{\gamma \delta }
+c^N_2(a^{\alpha \gamma }a^{\beta \delta }+a^{\alpha \delta }a^{\beta 
\gamma }),
\quad
q^{\alpha \beta }=q a^{\alpha \beta }.
\end{equation*}
The coefficients $c^N_1$, $c^N_2$, $\epsilon ^{33}_P$, and $q$ in the two-dimensional electric enthalpy can be expressed through the 3-D electroelastic moduli by means of \eqref{4.3}. The results are
\begin{align}
c^N_1(\zeta )&=c^E_{12}-(c^E_{13})^2/c^E_{33}+(e_{31}-c^E_{13}e_{33}/c^E_{33})^2/(\epsilon ^S_{33}+(e_{33})^2/c^E_{33}), \notag
\\
c^N_2(\zeta )&=\frac{1}{2}(c^E_{11}-c^E_{12}),\quad \epsilon ^{33}_P(\zeta )=\epsilon ^S_{33}+(e_{33})^2/c^E_{33}, \label{8.1}
\\
q(\zeta )&=(e_{31}-c^E_{13}e_{33}/c^E_{33})/(\epsilon ^S_{33}+(e_{33})^2/c^E_{33}), \notag
\end{align}
where Voigt's abbreviated index notation is used for the 3-D electroelastic moduli standing on the right-hand sides of \eqref{8.1} \citep{le1999vibrations}.

Substituting the above formulas into the 2-D action functional \eqref{6.14} and taking into account that $u_1$ and $u$ do not depend on $x^2$, we reduce it to 
\begin{multline}
\label{8.2}
J=\pi Rh\int_{t_0}^{t_1}\int_{-L}^{L} \{ \bar{\rho } (\dot{u}_1^2+\dot{u}^2)-[\bar{c}_1^N (u_{1,1}+\frac{u}{R})^2+2\bar{c}_2^N u_{1,1}^2 +2\bar{c}_2^N\frac{u^2}{R^2}
\\
+2h\bar{a}_1^N(u_{1,1}+\frac{u}{R})u_{,11}+4h\bar{a}_2^Nu_{1,1}u_{,11}+h^2(\bar{b}_1^N+2\bar{b}_2^N)u_{,11}^2
\\
-\bar{\epsilon }(-\frac{\varphi _0(t)}{h}+\bar{q}u_{1,1}+\bar{q}\frac{u}{R}-h\bar{p}u_{,11})^2] \} \, dx^1 \, dt,
\end{multline}
where the following short notations for the coefficients are used
\begin{gather}
\bar{c}^N_1=\langle c^N_1(\zeta ) \rangle, \quad \bar{a}^N_1=\langle c^N_1(\zeta )\zeta \rangle , \quad \bar{b}^N_1=\langle c^N_1(\zeta )\zeta ^2\rangle , \notag
\\
\bar{c}^N_2=\langle c^N_2(\zeta ) \rangle, \quad \bar{a}^N_2=\langle c^N_2(\zeta )\zeta \rangle , \quad \bar{b}^N_2=\langle c^N_2(\zeta )\zeta ^2\rangle , \label{8.2a}
\\
\bar{q}=\langle q(\zeta ) \rangle, \quad \bar{p}=\langle q(\zeta )\zeta \rangle, \quad \bar{\epsilon }=\langle \frac{1}{\epsilon ^{33}_P(\zeta )} \rangle ^{-1}, \quad \bar{\rho }=\langle \rho (\zeta )\rangle . \notag 
\end{gather}
Varying functional \eqref{8.2}, we obtain the Euler equation
\begin{equation}
\bar{\rho }\ddot{u}_1 =(\bar{c}^N_1-\bar{\epsilon }\bar{q}^2)(u_{1,11}+\frac{u_{,1}}{R})+2\bar{c}^N_2u_{1,11}+h(\bar{a}^N_1+2\bar{a}^N_2+\bar{\epsilon }\bar{p}\bar{q})u_{,111} ,
\label{8.3}
\end{equation}
and
\begin{multline}
\bar{\rho }\ddot{u}=-\frac{1}{R}(\bar{c}^N_1-\bar{\epsilon }\bar{q}^2)(u_{1,1}+\frac{u}{R})-2\bar{c}^N_2\frac{u}{R^2}-\frac{\bar{\epsilon }\bar{q}}{R}\frac{\varphi _0(t)}{h} 
\\
-h(\bar{a}^N_1+2\bar{a}^N_2+\bar{\epsilon}\bar{p}\bar{q})u_{1,111}-2h(\bar{a}^N_1+\bar{\epsilon}\bar{p}\bar{q})\frac{u_{,11}}{R}-h^2(\bar{b}^N_1+2\bar{b}^N_2-\bar{\epsilon}\bar{p}^2)u_{,1111}. \label{8.4}
\end{multline}
For the free edges of the shell we have the following boundary conditions at $x^1=\pm L$
\begin{equation}\label{8.5} 
\begin{split}
(\bar{c}^N_1-\bar{\epsilon }\bar{q}^2)(u_{1,1}+\frac{u}{R})+2\bar{c}^N_2u_{1,1}+h(\bar{a}^N_1+2\bar{a}^N_2+\bar{\epsilon}\bar{p}\bar{q})u_{,11}+\bar{\epsilon}\bar{q}\frac{\varphi _0(t)}{h}=0,
\\
h(\bar{b}^N_2+2\bar{b}^N_2-\bar{\epsilon}\bar{p}^2)u_{,11}+(\bar{a}^N_1+2\bar{a}^N_2+\bar{\epsilon}\bar{p}\bar{q})u_{1,1}+(\bar{a}^N_1+\bar{\epsilon}\bar{p}\bar{q})\frac{u}{R}-\bar{\epsilon}\bar{p}\frac{\varphi _0(t)}{h}=0,
\\
h(\bar{b}^N_2+2\bar{b}^N_2-\bar{\epsilon}\bar{p}^2)u_{,111}+(\bar{a}^N_1+2\bar{a}^N_2+\bar{\epsilon}\bar{p}\bar{q})u_{1,11}+(\bar{a}^N_1+\bar{\epsilon}\bar{p}\bar{q})\frac{u_{,1}}{R}=0.
\end{split}
\end{equation}
The voltage $\varphi _0(t)$ is assumed to depend harmonically on $t$, $\varphi _0(t)=\hat{\varphi }_0 \cos (\omega t)$, so that solutions of \eqref{8.3}, \eqref{8.4}, \eqref{8.5} can be sought in the form 
\begin{equation}
\label{8.6}
u_1(x^1,t)=\hat{u}_1(x^1)\cos(\omega t),\quad u(x^1,t)=\hat{u}(x^1)\cos(\omega t).
\end{equation}
Introducing the dimensionless variable and quantities
\begin{gather}
\zeta ^1=\frac{x^1}{R}, \quad \vartheta =\omega R \sqrt{\frac{\bar{\rho }}{\bar{c}^N_1+2\bar{c}^N_2-\bar{\epsilon }\bar{q}^2}}, \quad \mu _1=\frac{\bar{c}^N_1-\bar{\epsilon }\bar{q}^2}{\bar{c}^N_1+2\bar{c}^N_2-\bar{\epsilon }\bar{q}^2},
\\
\mu _2=\frac{\bar{a}^N_1+2\bar{a}^N_2+\bar{\epsilon }\bar{p}\bar{q}}{\bar{c}^N_1+2\bar{c}^N_2-\bar{\epsilon }\bar{q}^2}, \quad \mu _3=\frac{2(\bar{a}^N_1+\bar{\epsilon }\bar{p}\bar{q})}{\bar{c}^N_1+2\bar{c}^N_2-\bar{\epsilon }\bar{q}^2},\quad \mu _4=\frac{\bar{b}^N_1+2\bar{b}^N_2-\bar{\epsilon }\bar{p}^2}{\bar{c}^N_1+2\bar{c}^N_2-\bar{\epsilon }\bar{q}^2},
\\
h_* =\frac{h}{R}, \quad \beta =\frac{\bar{\epsilon }\bar{q}}{\bar{c}^N_1+2\bar{c}^N_2-\bar{\epsilon }\bar{q}^2}, \quad \nu =\frac{\bar{p}}{\bar{q}}
\end{gather}
this system can be transformed to the differential equations
\begin{equation}\label{8.8}
\begin{split}
\hat{u}^{\prime \prime }_1+\mu _1 \hat{u}^{\prime }+\mu _2 h_* \hat{u}^{\prime \prime \prime }+\vartheta ^2 \hat{u}_1=0,
\\
\mu _1 \hat{u}^{\prime }_1+\hat{u}+\mu _2h_*\hat{u}^{\prime \prime \prime }_1+\mu _3h_* \hat{u}^{\prime \prime }+\beta R \frac{\hat{\varphi }_0}{h}+\mu _4 h_*^2 \hat{u}^{\prime \prime \prime \prime }-\vartheta ^2 \hat{u}=0,
\end{split}
\end{equation}
with prime denoting the derivative with respect to $\zeta ^1$. The boundary conditions at $\zeta ^1=\pm l=\pm L/R$ become
\begin{equation}\label{8.9}
\begin{split}
\hat{u}^{\prime }_1+\mu _1 \hat{u}+\mu _2 h_* \hat{u}^{\prime \prime }+\beta R \frac{\hat{\varphi }_0}{h}=0,
\\
\mu _4h_* \hat{u}^{\prime \prime }+\mu _2\hat{u}^{\prime }_1+\mu _3 \hat{u}-\beta \nu R \frac{\hat{\varphi }_0}{h}=0,
\\
\mu _4 h_* \hat{u}^{\prime \prime \prime }+\mu _2\hat{u}^{\prime \prime }_1+\mu _3\hat{u}^\prime =0.
\end{split}
\end{equation}

It is easy to see that the symmetric solutions of \eqref{8.8} and \eqref{8.9} are given by
\begin{equation}
\hat{u}_1=\sum_{i=1}^3 a_i\sin \kappa _i\zeta ^1, \quad 
\bar{u}=-\frac{\beta R\hat{\varphi }_0}{h(1-\vartheta ^2)}
+\sum_{i=1}^3 a_i \gamma _i \cos \kappa _i\zeta ^1,
\label{8.10}
\end{equation}
where $\kappa _i^2$ are the roots of the cubic equation
\begin{equation}\label{8.10a}
(-\kappa ^2+\vartheta ^2)(\mu _4h_*^2 \kappa ^4-\mu _3h_*\kappa ^2+1 
-\vartheta ^2)+(\mu _2h_*\kappa ^2-\mu _1)^2\kappa ^2=0,
\end{equation}
$\gamma _i$ are expressed through $\kappa _i$ in the following way
\[
\gamma _i=\frac{\kappa _i^2-\vartheta ^2}{(\mu _2h_*\kappa _i^2-\mu _1)\kappa _i},
\]
and $a_i$ should be found as the solution of the system of linear 
equations
\begin{equation*}
\sum_{j=1}^3C_{ij}a_{j}=b_i, \quad i=1,2,3.
\end{equation*}
The elements of the $3\times 3$ matrix $C_{ij}$ are equal to
\begin{align}
C_{1j}&=(\kappa _j+\mu _1\gamma _j-\mu _2\kappa _j^2\gamma _j)\cos \kappa _jl, \notag
\\
C_{2j}&=(-\mu _4h_* \kappa _j^2 \gamma _j+\mu _2\kappa _j+\mu _3 \gamma _j)\cos \kappa _jl, 
\label{8.11}
\\
C_{3j}&=(\mu _4h_* \kappa _j^3 \gamma _j-\mu _2\kappa _j^2-\mu _3 \kappa _j \gamma _j)\sin \kappa _jl ,
\end{align}
while $b_i$ are given by
\begin{equation*}
b_1=\beta \frac{R\hat{\varphi }_0}{h}(\frac{\mu _1}{1-\vartheta ^2}-1), \quad b_2=\beta \frac{R\hat{\varphi }_0}{h}(\frac{\mu _3}{1-\vartheta ^2}+\nu ), \quad b_3=0.
\end{equation*}

After finding $a_i$ we determine the amplitude of $D^3$ by \eqref{6.18} yielding
\begin{multline*}
\hat{D}^3=\bar{\epsilon }[-\frac{\hat{\varphi }_0}{h}
+\bar{q}(\hat{u}_{1,1}+\frac{\hat{u}}{R})-h\bar{p}\bar{u}_{,11}]=-\bar{\epsilon }\frac{\hat{\varphi }_0}{h}-\frac{\bar{\epsilon }\bar{q}\beta \hat{\varphi }_0}{h(1-\vartheta ^2)}
\\
+\frac{\bar{\epsilon }\bar{q}}{R}\sum_{i=1}^3 a_i (\kappa _i+\gamma _i )\cos \kappa _i\zeta ^1+
\frac{\bar{\epsilon }\bar{q}}{R}\sum_{i=1}^3 \nu h_* a_i \kappa _i ^2\gamma _i\cos \kappa _i\zeta ^1.
\end{multline*}
Then the amplitude of the total charge on one of the electrodes is equal to
\begin{equation*}
\int_{\Omega_+}\hat{D}^3\, da=4\pi R^2\bar{\epsilon } \frac{\hat{\varphi }_0}{h} [ -(1+\frac{\bar{q}\beta }{1-\vartheta ^2})l+\bar{q}\beta \sum_{i=1}^3\bar{a}_i(1+\frac{\gamma _i}{\kappa _i}+\nu h_*\kappa _i\gamma _i)\sin \kappa _i l]
\end{equation*}
where $\bar{a}_i$ is the solution of the system
\begin{equation}\label{8.11a}
\sum_{j=1}^3C_{ij}\bar{a}_j=\bar{b}_i,\quad i=1,2,3,
\end{equation}
with 
\[
\bar{b}_1=\frac{\mu _1}{1-\vartheta ^2}-1,\quad \bar{b}_2=\frac{\mu _3}{1-\vartheta ^2}+\nu  ,\quad \bar{b}_3=0.
\]

\begin{table}[htb]
  \centering
\begin{tabular}{|l|c|c|c|c|c|c|c|c|c|c|c|c|} \hline
 $c_{11}^E$ & $c^E_{12}$ &  $c^E_{13}$ &
  $c^E_{33}$ & $c^E_{44}$ & $e_{15}$ & $e_{31}$ & $e_{33}$ & $\epsilon ^S_{11}$ & $\epsilon ^S_{33}$ & $\rho $ \\ \hline
13.9 & 7.8 & 7.4 & 11.5 & 2.56 & 12.7 & -5.2 & 15.1 & 650 & 560 & 7500 \\ \hline
20.97 & 12.11 & 10.51 & 21.09 & 4.25 & -0.59 & -0.61 & 1.14 & 7.38 & 7.83 & 5676 \\ \hline
\end{tabular}
  \caption{Material constants of PZT-4 (first row) and ZnO (second row)}
  \label{table1}
\end{table}

According to the solution of this problem the resonant frequencies are the roots of the determinantal equation
\begin{equation}
\det C_{ij} =0.
\label{8.12}
\end{equation}
The antiresonant frequencies should be found from the condition that the total charge vanishes giving
\begin{equation}
\bar{q}\beta \sum_{i=1}^3\bar{a}_i(1+\frac{\gamma _i}{\kappa _i}+\nu h_*\kappa _i\gamma _i)\sin \kappa _i l = (1+\frac{\bar{q}\beta }{1-\vartheta ^2})l.
\label{8.13}
\end{equation}

For the numerical simulations we use as an example an FG piezoceramic material, whose 3-D electroelastic moduli vary in the thickness direction according to \citep{reddy1999axisymmetric}
\begin{equation}
\label{8.14}
M(\zeta )=M_{PZT}(1/2-\zeta )^\lambda +M_{ZnO}[1-(1/2-\zeta )^\lambda ],
\end{equation}
where $\lambda $ is the gradient index, $M$ indicates any component of 3-D electroelastic moduli, while $M_{PZT}$ and $M_{ZnO}$ are the corresponding material constants of PZT-4 and ZnO presented in Table 1 (with the unit for $c^E_{\mathfrak{ab}}$ being $10^{10}$N/m$^2$, $e_{i\mathfrak{a}}$ - C/m$^2$, $\epsilon ^S_{ij}$ - $10^{-11}$F/m, and $\rho $ - kg/m$^3$) \citep{chen2002free}.

\begin{figure}[htb]
    \begin{center}
    \includegraphics[width=7cm]{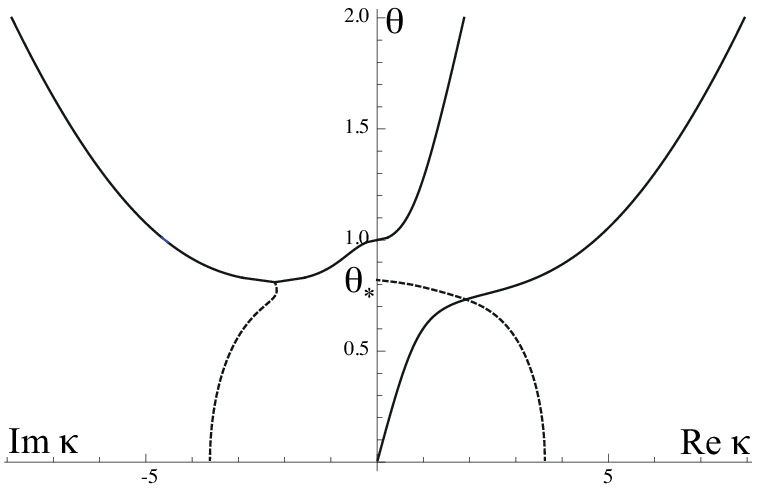}
    \end{center}
    \caption{Dispersion curve corresponding to equation (66).}
    \label{fig:4}
\end{figure}

\begin{figure}[htb]
    \begin{center}
    \includegraphics[width=7cm]{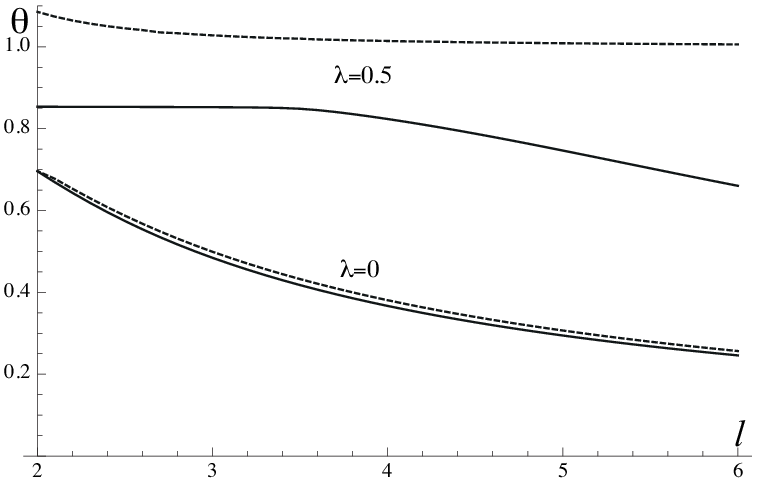}
    \end{center}
    \caption{First resonant (bold line) and anti-resonant (dashed line) frequencies versus $l$.}
    \label{fig:5}
\end{figure}

With these material data we compute the numerical values of constants used in this 2-D problem in accordance with \eqref{8.1}, \eqref{8.14}, and \eqref{8.2a}. The results of numerical simulations are shown in Figures 4 and 5, where we took $h_*=0.1$. Figure 4 plots the typical dispersion curves (for $\lambda =1$) which correspond to three different roots $\kappa _1$, $\kappa _2$, $\kappa _3$ (up to their sign) of equation \eqref{8.10a} for each $\vartheta $. We observe that in $(0,\vartheta _*)$ there are two complex conjugate roots and one real root, in $(\vartheta _*,1)$ two imaginary roots and one real root, and in the remaining region of $\vartheta $ two real roots and one imaginary root. The dash lines in Figure 4 correspond to the real and imaginary parts of the same curve in the complex plane of $\kappa $. Figure 5 shows the first resonant and anti-resonant frequencies of axisymmetric vibrations as functions of the half-length to radius ratio $l=L/R$ for $\lambda =0$ (homogeneous PZT-4 piezoceramic material) and $\lambda =0.5$ (FGP-material). To minimize the numerical errors in solving the equation \eqref{8.11a} and computing the determinant \eqref{8.12}, the sine and cosine functions in the matrix $C_{ij}$ should be normalized by $e^{d_i l}$, with $d_i$ being the positive real parts of the roots. It is interesting to observe the strong difference in behavior of the resonant and anti-resonant frequencies of the FGP-shell as compared to those of the homogeneous piezoceramic shells. For FGP-shell the anti-resonant frequency is nearly insensitive to the change of $l$ and lies much higher above the resonant one. Note also that for $\lambda =0$ these frequencies coincide with those found by \citet{le1999vibrations}.

\section{Conclusion}
It is shown in this paper that the rigorous first order approximate 2-D theory of thin FGP-shells can be derived from the exact 3-D piezoelectricity theory by the variational-asymptotic method. The electroelastic fields of the FGP-shell vary through the thickness and differ essentially from those of the homogeneous piezoelectric shell. The error estimation for the constructed 2-D theory is established that enables one to apply this theory to the vibration control of thin FGP-shells.

\end{document}